\documentclass[english,aps,prb,twocolumn,nofootinbib,showkeys,longbibliography,secnumarabic,superscriptaddress,floatfix]{revtex4-1}

\usepackage[english]{babel}
\usepackage{amssymb,amsmath,soul}

\usepackage[pdftex]{graphicx}
\usepackage[caption=false,listofformat=parens]{subfig}

\usepackage{algorithmicx}
\usepackage[Algorithm,ruled]{algorithm}

\usepackage{algpseudocode}

\usepackage{color}
\usepackage[usenames,dvipsnames]{xcolor}
\usepackage{hyperref} 
\hypersetup{%
	linktoc=page,%
	colorlinks=true,%
	linkcolor=RoyalBlue,%
	citecolor=OliveGreen,%
	urlcolor=OliveGreen,%
	linkbordercolor=red,%
	pdfborderstyle={/S/U/W 1},%
}
\setul{1.4pt}{.4pt}

\usepackage{booktabs,tabularx}
\newcolumntype{R}{>{\raggedleft\arraybackslash}X}

\usepackage{enumerate}
\usepackage{grffile}
\usepackage{balance}
\usepackage{float}

\newcommand{\Jc}{J_\mathrm{c}}

\newcommand{\Jext}{J}
\newcommand{\Tc}{T_\mathrm{c}}
\newcommand{\Ec}{E_\mathrm{c}}
\newcommand{\Hct}{H_\mathrm{c2}}
\newcommand{\fL}{\mathbf{f}_\mathrm{\scriptscriptstyle L}} 
\newcommand{\x}{\mathbf{x}}
\newcommand{\xopt}{\mathbf{x}_\mathrm{opt}}
\newcommand{\p}{\mathbf{p}}
\newcommand{\q}{\mathbf{q}}
\newcommand{\qopt}{\mathbf{q}_\mathrm{opt}}
\newcommand{\fopt}{f_\mathrm{opt}}
\newcommand{\f}{\rho}
\newcommand{\fs}{\rho_\mathrm{s}}
\newcommand{\Ds}{D_\mathrm{s}}
\newcommand{\Ns}{N_\mathrm{s}}
\newcommand{\fn}{\rho_\mathrm{n}}
\newcommand{\fc}{\rho_\mathrm{c}}
\newcommand{\Dc}{D_\mathrm{c}}
\newcommand{\Nc}{N_\mathrm{c}}
\newcommand{\fe}{\rho_\mathrm{e}}
\newcommand{\Dz}{D_{z}}
\newcommand{\tDz}{\tilde D_{z}}
\newcommand{\Dxy}{D_{xy}}
\newcommand{\Ne}{N_\mathrm{e}}

\begin{document}

\title{\textit{In silico} optimization of critical currents in superconductors}

\author{Gregory Kimmel}
\affiliation{Materials Science Division, Argonne National Laboratory, 9700 S. Cass Av., Argonne, IL 60637, USA}
\affiliation{Department of Engineering Sciences and Applied Mathematics, Northwestern University, 633 Clark St, Evanston, IL 60208, USA}

\author{Ivan A. Sadovskyy} 
\affiliation{Materials Science Division, Argonne National Laboratory, 9700 S. Cass Av., Argonne, IL 60637, USA}
\affiliation{Computation Institute, University of Chicago, 5735 S. Ellis Av., Chicago, IL 60637, USA}

\author{Andreas Glatz}
\affiliation{Materials Science Division, Argonne National Laboratory, 9700 S. Cass Av., Argonne, IL 60637, USA}
\affiliation{Department of Physics, Northern Illinois University, DeKalb, IL 60115, USA}

\date{\today}

\begin{abstract}
For many technological applications of superconductors the performance of a material is determined by the highest current it can carry losslessly~--- the critical current. In turn, the critical current can be controlled by adding non-superconducting defects in the superconductor matrix. Here we report on systematic comparison of different local and global optimization strategies to predict \textit{optimal} structures of pinning centers leading to the highest possible critical currents. We demonstrate performance of these methods for a superconductor with randomly placed spherical, elliptical, and columnar defects. 
\end{abstract}

\maketitle


\section{Introduction}

The most important feature of superconductors for high-performance applications~\cite{Kwok:2016} is its ability to carry large currents with almost no dissipative loss. At the same time, recent advances have made it possible to manufacture {type-II}-superconductor-based cables on industrial scales making large-scale application possible.\cite{Holesinger:2008,Malozemoff:2012} The main dissipative mechanism in these type-II superconductors stems from the motion of magnetic vortices within the superconducting matrix. These are normal elastic filaments, carrying quantized magnetic fluxes,\cite{Abrikosov:1957} which appear in these superconductors in magnetic fields larger than the first critical field~--- in contrast to {type-I} superconductors, which lose their superconducting properties once the magnetic field enters the material. The highest amount of current that can be passed through a superconductor without dissipation is known as the critical current, which strongly depends on the vortex dynamics and its interaction with non-superconducting defects. Strictly speaking, the critical current in the presence of vortices is always zero at finite temperatures due to thermal creep, but it is conventionally defined as the current at which the dissipation/voltage reaches some threshold value. At sufficiently large currents the dissipation and associated heating of the superconductor will ultimately lead to the loss of superconductivity.

Despite all technological advances, typical critical currents are still way below the theoretical limit and from an economic point of view, too low to be competitive with conventional cables for large-distance energy transport. An obvious way to improve the critical current in existing superconductors is to impede the vortex motion more effectively. This can be accomplished by placing defects, which can ``pin'' magnetic vortices and thus prevent their motion,\cite{Blatter:1994,Blatter:2003} in a sophisticated way. The entirety of all defects is also called the pinning landscape (or \textit{pinscape}), which in conjunction with intrinsic material inhomogeneities and the sample geometry define the critical current of a given sample.

This defines the task, which we consider in this work: the optimization of the pinscape for highest possible critical currents. More precisely, we concentrate on defects or pinning centers, which can be controlled during the fabrication of a sample, like self-assembled inclusions\cite{Ortalan:2009,Sadovskyy:2016a} or irradiation defects.\cite{Sadovskyy:2016b,Leroux:2015,Eley:2017} The efficiency of the pinscape strongly depends on the shape, size, and arrangement of individual pinning centers. Indeed, bigger defects ensure a larger pinning force, but, at the same time, reduce the effective cross-section of the superconductor needed for current flow. The optimal pinscape also depends on the intended application, particularly on the type of superconductor and on the value and the direction of external magnetic fields. The question we address in this paper is, how one can find the best pinscape most effectively. This systematic prediction of optimal pinscapes for a given application aims at replacing the traditional trial-and-error approach.\cite{Sadovskyy:2016b}

Here, we test several optimization strategies needed for a systematic improvement of the critical current in superconductors. We compare the efficiency of a global method (particle swarm optimization) and three local methods (Nelder-Mead method, pattern search, and adaptive pattern search) for a typical critical current optimization problem. The critical current for a given pinscape is calculated using a GPU-based iterative solver for the time-dependent Ginzburg-Landau (TDGL) equation describing {type-II} superconductors.\cite{Sadovskyy:2015a} This model correctly captures the vortex dynamics\cite{Blatter:1994,Blatter:2003,Vodolazov:2013} in superconductors in the vicinity of the critical temperature and is capable of reproducing experimental critical currents for a given pinscape.\cite{Berdiyorov:2006,Berdiyorov:2006b,Sadovskyy:2016a,Sadovskyy:2016b,Sadovskyy:2017} In addition, we provide a detailed analysis of these methods applied to several benchmark functions for comparison. 

The article is organized as follows. In Sec.~\ref{sec:methods} we formulate the general optimization problem and describe the optimization methods studied here. In Sec.~\ref{sec:benchmark_functions} we present a detailed comparison of the efficiency of the chosen optimization strategies on benchmark functions. In Sec.~\ref{sec:GL} we briefly describe the TDGL model for superconductors and define three  physically relevant optimization problems in Sec.~\ref{sec:problem}. The results of the optimization for these problems are discussed in Sec.~\ref{sec:results}. Finally, we summarize our results in Sec.~\ref{sec:conclusions}.

\section{Optimization methods and problem formulation} \label{sec:methods}

Optimization methods are divided into two classes: local and global methods. Examples of global search methods include particle swarm optimization (PSO), cuckoo search, simulated annealing, etc. Here we focus on PSO because of its straight forward parallelizability. We compare this global PSO method to three local methods: pattern search (aka coordinate descent), adaptive pattern search, and the Nelder-Mead method (aka downhill simplex method). The pattern search and Nelder-Mead methods are standard local methods and their analysis, convergence properties, and pitfalls have been widely studied.\cite{Hooke:1961,Torczon:1997,Audet:2002,Dolan:2003,Lagarias:1998,McKinnon:1998,Rios:2013} The adaptive pattern search method is a recent improvement on the traditional pattern search.\cite{Loshchilov:2011} These methods can be used in conjunction with some more sophisticated methods, or routines, such as multi-level single linkage, which begins with many different initial starting points, and collects them into multiple sets, which depend on whether they are sufficiently close to a previously found local optimum. If they are not close to any previously found optimum, a local search is started.\cite{RinnooyKan:1987a,RinnooyKan:1987b} The primary use behind this type of routine is to terminate local searches, which are falling into the basin of attraction of an optimum point already found, which reduces the computation time and number of function evaluations.

The general optimization problem of a function $f$ can be formulated as follows
\begin{equation}
	\xopt = \arg \, \min\limits_{\x \in \Omega} \{ f(\x) \}, \quad
	\fopt = \min\limits_{\x \in \Omega} \{ f(\x) \},
	\label{eq:opt} 
\end{equation}
where $\x \in \Omega$ is a set of parameters in parameter space~$\Omega$. $\x$ defines all variable properties in the system, which in turn, defines the objective function $f(\x)$.
In the case of a superconductor, the parameter set $\x$ can, e.g., determine the pinscape of the system and the corresponding objective function is $f(\x) = -\Jc(\x)$, where $\Jc(\x)$ is the critical current of the superconductor with pinscape $\x$. Note, that the minimization of this $f(\x)$ is equivalent to the maximization of $\Jc(\x)$.

\subsection{Particle swarm optimization}

The PSO algorithm is a meta-heuristic global optimization algorithm.\cite{Kennedy:2011,Poli:2007} Its convergence properties have been studied in a simplified form, where a single particle was used and the randomness in the algorithm was replaced by its averaged value.\cite{Trelea:2003} It performs well on all test problems, but is typically outshone by local methods when there was only a single minimum. The utility of this method is best seen on functions with multiple local optima, see Fig.~\subref{fig:Rastrigin_function}. This test function is relevant for problems having multiple local extrema, such as the problem of periodically arranged pinning centers in superconductors. In that case, many local maxima of~$\Jc$ exist at integer values of the ratio of number of pinning centers to number of vortices.\cite{Sadovskyy:2017}

The PSO has four main control parameters given by $\q = \{S$, $\omega$, $\phi_\mathrm{p}$, $\phi_\mathrm{g} \}$, where $S$ is the swarm size, $\omega$ the inertia of the individual particle (its tendency to move in its current direction), $\phi_\mathrm{p}$ and $\phi_\mathrm{g}$ are the weights for the particle to move towards its personal and global best point in parameter space, respectively. The pseudocode for PSO is presented in Listing~\ref{alg:particle_swarm_optimization}.

\begin{algorithm}[H]
\floatname{algorithm}{Listing}
\caption{Particle swarm optimization} \label{alg:particle_swarm_optimization}
\begin{algorithmic}[1]
	\State \textbf{input:} Lower and upper limits, $\mathbf{L}$ and $\mathbf{U}$
	\State \textbf{input:} Parameters $S$, $\omega$, $\phi_\mathrm{p}$, and $\phi_\mathrm{g}$
	\State \textbf{input:} Parameters $K_\mathrm{exit}$ from $[1, S]$ and $D_\mathrm{exit}$
	\For {$i = 1$, $\dots$, $S$}
		\State Uniformly distributed particle position $\x_i \gets U_n(\mathbf{L}, \mathbf{U})$
		\State Particle velocity $\mathbf{v}_i \gets U_n(\mathbf{L} -\mathbf{U}, \mathbf{U} -\mathbf{L})$
		\State Particle best known position $\p_i \gets \x_i$
	\EndFor
	\State Best global position $\mathbf{g} = \mathrm{arg} \min\limits_{i} f(\p_i)$
	\Repeat
		\For {$i = 1$, $\dots$, $S$}
			\State $\mathbf{r}_\mathrm{p} \gets U_n(0,1)$, \quad
			$\mathbf{r}_\mathrm{g} \gets U_n(0,1)$
			\State $\mathbf{v}_i \gets \omega \mathbf{v}_i + \phi_\mathrm{p}  \mathbf{r}_\mathrm{p}( \p_i - \x_i) + \phi_\mathrm{g} \mathbf{r}_\mathrm{g} (\mathbf{g} - \x_i)$
			\State $\x_i \gets \x_i + \mathbf{v}_i$
			\If {$f(\x_i) < f(\p_i)$}
				\State $\p_i \gets \x_i$
				\If {$f(\p_i) < f(\mathbf{g})$}
				\State $\mathbf{g} \gets \p_i$
				\EndIf
			\EndIf
			\State Distance from global best position $d_i \gets \Bigl\| \cfrac{\x_i - \mathbf{g}}{\mathbf{U} -\mathbf{L}} \Bigr\|_2$ \label{alg:PSO_exit_criterion_d}
		\EndFor
		\State Sort $d_1$, $d_2$, $\ldots$, $d_S$ in ascending order
		\State ${\bar d} \gets \cfrac{1}{K_\mathrm{exit}} \sum\limits_{i=1}^{K_\mathrm{exit}} d_i$ \label{alg:PSO_exit_criterion_avd}
	\Until {${\bar d} < D_\mathrm{exit}$} \label{alg:PSO_exit_criterion}
	\State \textbf{output:} $\mathbf{g}$
\end{algorithmic}
\end{algorithm}

The objective function is updated independently for each particle, which makes PSO parallelizable. In this way, a large architecture can make this method highly efficient for multimodal surfaces, where a large swarm size can be way more efficient in converging towards a global solution then local methods. 

The biggest challenge is in determining an appropriate exit criterion for the routine. We use 

\begin{enumerate}
\item[(i)] the change, $|f(\mathbf{g}_\mathrm{best}) - \langle f(\mathbf{g}) \rangle_{\scriptscriptstyle M}|$, in the best found objective function value, $f(\mathbf{g}_\mathrm{best})$, with $f(\mathbf{g})$ averaged over the last $M$ iterations of PSO, $\langle f(\mathbf{g}) \rangle_{\scriptscriptstyle M}$; or 

\item[(ii)] the average distance of the particle from the swarm global best position $K_\mathrm{exit}^{-1} \sum_{i=1}^{K_\mathrm{exit}} \| (\x_i - \mathbf{g}) / (\mathbf{U} -\mathbf{L}) \|_2$, where the sum is taken over $K_\mathrm{exit}$ particles with the best objective values, see lines~\ref{alg:PSO_exit_criterion_d}, \ref{alg:PSO_exit_criterion_avd}, and \ref{alg:PSO_exit_criterion} in Listing~\ref{alg:particle_swarm_optimization}. A small variant would be to use $\p_i$ instead $\x_i$ of each respective particle, which would be less sensitive to the exploratory nature of PSO (through the inertial term $\omega \mathbf{v}_i$).
\end{enumerate}

The first exit criterion avoids taking an unnecessary amount of evaluations if there is no sufficient improvement (the method could then be terminated, switched to a local method, or restarted). The second was done to avoid cases where there may be many local optimum and so a few particles get trapped among them. This avoid premature stopping if many particles are spread out in parameter space and the improvement in $\mathbf{g}$ is slow. We stop when a certain proportion  of the swarm has been attracted to the best found point. A higher proportion means more evaluations but also a higher probability that we converged to the correct optimum. Therefore, there is a balance between speed and accuracy. In the testing on the benchmark functions, we found no significant difference in accuracy for $K_\mathrm{exit} > S/2$, but a difference in speed. In the simulations presented below we chose $K_\mathrm{exit} = 0.7 S$.

\subsection{Pattern search}

Pattern search is a straightforward method, which starting from a random point, evaluates $2n + 1$ points (including the initial point) where $n$ is the dimension of the parameter space, by moving a distance along each dimension in the search space. It then moves to the point which improves the function the most. The method then evaluates $2n - 1$ new points (it does not need to re-evaluate the point its on or the point it came from). If no improvement is made, the step size is reduced. Once the step size is below some threshold, it exits out of the loop. Along with its simplicity comes its ability to converge to non-stationary problems on some relatively simple problems.\cite{Powell:1973} It has a particularly difficult time with functions with coordinate systems which are highly correlated, e.g., roughly speaking, the function gradient is not along any of the main coordinate axes, such as the Rosenbrock function shown in Fig.~\subref{fig:Rosenbrock_function}.

\subsection{Adaptive pattern search} \label{sec:adaptive_pattern_search}

The adaptive pattern search method is a recent modification of pattern search. The algorithm works similar to pattern search, however, while searching the parameter space, it adapts the coordinate system to achieve faster convergence, see Listing~\ref{alg:adaptive_encoding} adapted from Ref.~\onlinecite{Loshchilov:2011}. The presented adaptive encoding is similar to principal component analysis (PCA). The covariance matrix $\mathbf{C}$ and the transformation matrix $\mathbf{B}$ are updated using the most successful $\mu$ points. The transformation matrix modifies the coordinates to make them as uncorrelated as possible. Unlike PCA which typically looks to reduce dimensionality by retaining only eigenvectors corresponding to the largest eigenvalues, adaptive encoding retains all components.

\begin{algorithm}[H]
\floatname{algorithm}{Listing}
\caption{Adaptive encoding} \label{alg:adaptive_encoding}
\begin{algorithmic}[1]
	\State \textbf{input:} Parameters $\mu$, $\sigma$, $k_\mathrm{s}$, and $k_\mathrm{u}$
	\State \textbf{input:} $\mu$ best points $\x_1$, $\ldots$, $\x_\mu$
	\If {Initialize}
		\State $w_i \gets \cfrac{1}{\mu}$,\quad
			$c_p \gets \cfrac{1}{\sqrt n}$,\quad
			$c_1 \gets \cfrac{1}{2n}$,\: and\:
			$c_\mu \gets \cfrac{1}{2n}$
		\State $\p \gets 0$
		\State $\mathbf{C} \gets \mathbf{I};\quad \mathbf{B} \gets \mathbf{I}$
		\State $m \gets \sum\limits_{i=1}^\mu x_i w_i$
	\Else
	\State $m^- \gets m$
	\State $m \gets \sum\limits_{i=1}^\mu x_i w_i$
	\State $z_0 \gets \cfrac{\sqrt n}{\|\mathbf{B}^{-1}(m-m^-) \|} \, (m- m^-)$
	\For {$i = 1$, $\dots$, $\mu$}
		\State $z_i \gets \cfrac{\sqrt n}{\|\mathbf{B}^{-1}(x_i-m^-) \|} \, (x_i- m^-)$
	\EndFor
	\State $\p \gets (1-c_p) \p + \sqrt{c_p(2-c_p)} z_0$
	\State $\mathbf{C}_\mu \gets \sum\limits_{i=1}^\mu w_i z_i z_i^\mathrm{\scriptscriptstyle T}$
	\State $\mathbf{C} \gets (1- c_1 - c_\mu) \mathbf{C} + c_1 \p \p^\mathrm{\scriptscriptstyle T} + c_\mu \mathbf{C}_\mu$
	\State $\mathbf{B}^\mathrm{\scriptscriptstyle H} \mathbf{D} \mathbf{D} \mathbf{B}^\mathrm{\scriptscriptstyle H} \gets \mathtt{eigendecomposition}(\mathbf{C})$
	\State $\mathbf{B} \gets \mathbf{B}^\mathrm{\scriptscriptstyle H} \mathbf{D}$
	\EndIf
	\State \textbf{output:} $\p$, $\mathbf{B}$, $\mathbf{C}$
\end{algorithmic}
\end{algorithm}

Apart from the adaptive encoding, adaptive pattern search is very similar to pattern search. Suppose, we have an $n$-dimensional optimization problem, it initially searches in each coordinate direction, independently recording the new optimal as it progresses (in contrast to the generic pattern search we employed, which is not done sequentially). Then it keeps the best $\mu$ points, $\mu < 2n + 1$, and Listing~\ref{alg:adaptive_encoding} is called. This transformation is updated after each sweep of all $n$ dimensions ($2n - 1$ points) and applied in the next iteration. There are four parameters $\q = \{ \mu$, $\sigma$, $k_\mathrm{s}$, $k_\mathrm{u} \}$ which control the success and efficiency of this method: $\mu$ is the number of points used in the adaptive encoding function call, $\sigma$ is the initial step size, and $k_\mathrm{s}$ ($k_\mathrm{u}$) are the increase (decrease) of the step size upon successful (unsuccessful) improvement of the function value. The convergence rate was observed to be most sensitive to the parameter $\mu$. The utility of this method is best seen when applied to the Rosenbrock function, where pattern search needs over 20,000 iterations to converge in a two-dimensional parameter space, while the adaptive pattern search method typically converged in less than 1,000 iterations.

The key for the method's improved performance on Rosenbrock-type functions, is the adaptive encoding part of the algorithm. At this point we remark that the adaptive pattern search is typically a more efficient/performing method compared to the generic pattern search, but comparison to the latter offers a useful way to measure the overall shape of the optimal solution. We can logically deduce from the ratio of performance between these two methods, that if adaptive pattern search is orders of magnitude better than pattern search, then this implies that the parameters of interest are highly correlated.

\subsection{Nelder-Mead method}

The Nelder-Mead method (or downhill simplex) was chosen for its relative simple structure and its independence on the choice of the chosen coordinate system (it does not move along each dimension sequentially). This is most easily seen in comparing the method against the Rosenbrock function. This method utilizes simplexes, which are polytopes with $n + 1$ points (or vertices) in $n$ dimensions (i.e., triangles for $n = 2$, or tetrahedrons in $n = 3$).

Over time, many variants of Nelder-Mead have been conceived.\cite{Price:2002,Nazareth:2002,Conn:2009} Following Ref.~\onlinecite{Conn:2009}, Listing~\ref{alg:Nelder_Mead_method} describes the general Nelder-Mead method. Here we used standard values for parameters $\alpha = 1$, $\gamma = 2$, $\rho = 1/2$, and $\sigma = 1/2$. The exit criteria in line~\ref{alg:Nelder_Mead_exit_criteria} is satisfied when the points are within a certain distance and is defined by the ``volume'' 
\begin{equation*}
	\mathcal{V}(\x_1,\; \ldots,\; \x_{n+1}) 
	= \frac{|\text{det}(\mathbf{V})|}{n! \, \Delta^n},
\end{equation*}
where $\Delta = \min \|\x_i - \x_j\|$ and matrix $\mathbf{V} = (\mathbf{\dot x}_1$, $\ldots$, $\mathbf{\dot x}_n)$ consists of $n$ vectors $\mathbf{\dot x}_i = \x_{i+1} - \x_1$. This choice of the exit criteria was justified in Ref.~\onlinecite{Conn:2009}.

\begin{algorithm}[H]
\floatname{algorithm}{Listing}
\caption{Nelder-Mead method} \label{alg:Nelder_Mead_method}
\begin{algorithmic}[1]
\State \textbf{input:} Lower and upper limits, $\mathbf{L}$ and $\mathbf{U}$
\State \textbf{input:} Parameters $\alpha$, $\gamma$, $\rho$, and $\sigma$
\State Vertices $\x_1, \dots \x_{n+1} \gets U_{n+1}(\mathbf{L}, \mathbf{U})$
\Repeat
	\State Order the vertices $f(\x_1) \leqslant f(\x_2) \leqslant \ldots \leqslant f(\x_{n+1})$
	\State Centroid of $n$ best points $\x_0 \gets \cfrac{1}{n} \sum\limits_{i=1}^n \x_i$
	\State Reflected point $\x_\mathrm{r} \gets \x_0 + \alpha(\x_0 - \x_{n+1})$
	\If {$f(\x_1) \leqslant f(\x_\mathrm{r}) < f(\x_n)$}
		\State $\x_{n+1} \gets \x_\mathrm{r}$ 
	\ElsIf {$f(\x_\mathrm{r}) < f(\x_1)$,}
		\State Expanded point $\x_\mathrm{e} \gets \x_0 + \gamma(\x_\mathrm{r} - \x_0)$
		\If {$f(\x_\mathrm{e}) < f(\x_\mathrm{r})$}
			\State $\x_{n+1} \gets \x_\mathrm{e}$ 
		\Else
			\State $\x_{n+1} \gets \x_\mathrm{r}$ 
		\EndIf
	\Else
		\State Contracted point $\x_\mathrm{c} \gets \x_0 + \rho(\x_{n+1} - \x_0)$
		\If {$f(\x_\mathrm{c}) < f(\x_{n+1})$}
			\State $\x_{n+1} \gets \x_\mathrm{c}$ 
		\Else
			\State $\x_i \gets \x_1 + \sigma(\x_i - \x_1)$ for all $i > 1$
		\EndIf
	\EndIf
\Until {$\mathcal{V}(\x_1$, $\ldots$, $\x_{n+1}) < \epsilon$} \label{alg:Nelder_Mead_exit_criteria}
\State \textbf{output:} $\x_1$
\end{algorithmic}
\end{algorithm}

\section{Testing on benchmark functions} \label{sec:benchmark_functions}

\begin{figure*}
	\centering
	\vspace{-4mm}
	\subfloat{\includegraphics[width=14cm]{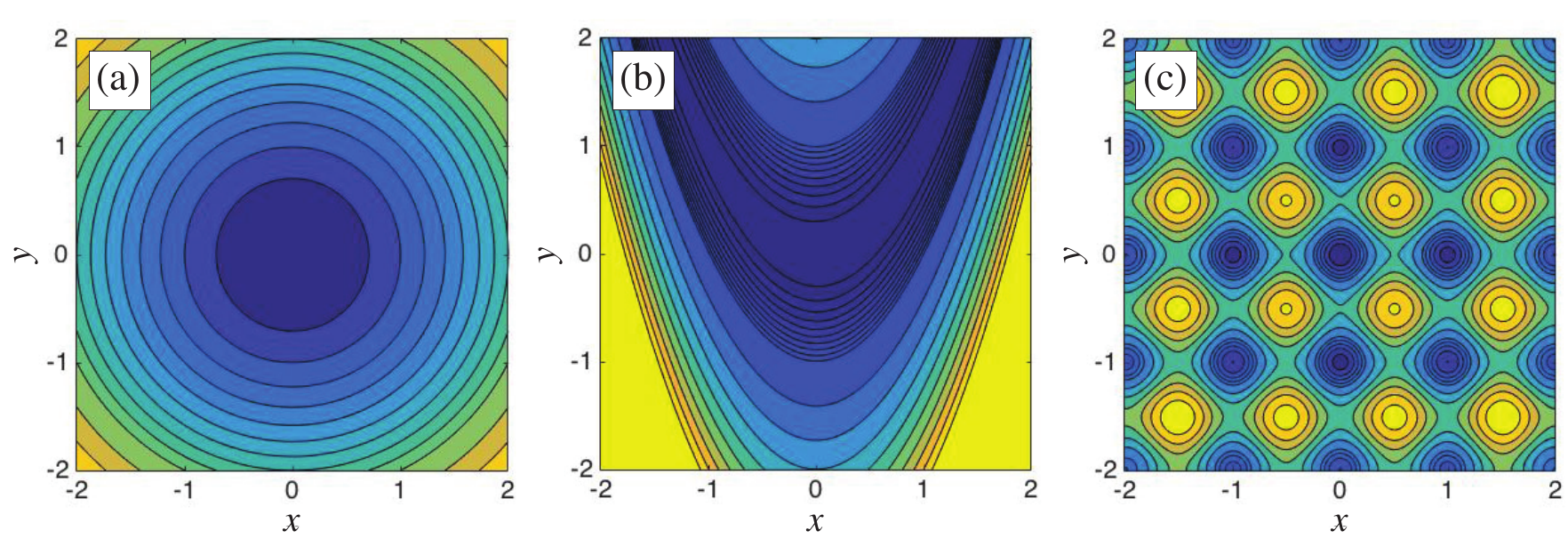} \label{fig:sphere_function}}
	\subfloat{\label{fig:Rosenbrock_function}}
	\subfloat{\label{fig:Rastrigin_function}}
	\vspace{-3mm}
	\caption{
		Examples of different test function with one global minimum.
		(a)~Sphere function, a simple multidimensional parabola.
		(b)~Rosenbrock function, a quartic multidimensional polynomial with shallow valley not along one of the coordinate axes.
		(c)~Rastrigin function, a periodic function with many local extrema.
	}
	\label{fig:benchmark_functions}
\end{figure*}

The above mentioned optimization methods are tested on the three benchmark functions shown in Fig.~\ref{fig:benchmark_functions}: the sphere function, the Rosenbrock function, and the Rastrigin function. These were chosen for their similarity to previously obtained critical-current surfaces in superconductors for low-dimensional sets of parameters. We expect scenarios where either a single optimum or multiple local with one global extrema exists. As a particular and practical example, the behavior of the solvers on the Rosenbrock function is related to the optimization of a pinscape consisting of spherical defects with two parameters being the number and diameter of the defects. In this case, it turns out that the Rosenbrock-type $\Jc$ surface can be removed by replacing the number of defects by the volume fraction. In general, however, one cannot be sure that an appropriate transformation exists or can even be found.

The study was broken up as follows: optimal parameters for the PSO and adaptive pattern search are obtained using the three benchmark functions. These were found by overlaying each respective optimization routine with PSO. For example, adaptive pattern search has tuning parameters $\q = \{ \mu$, $k_\mathrm{s}$, $k_\mathrm{u}$, $\sigma \}$, see Sec.~\ref{sec:adaptive_pattern_search}. The nested PSO algorithm then searches through this parameter space in an attempt to find the optimal parameters for the algorithm. We initially considered the function
\begin{equation} \label{eq:E_i}
	\bar E_{f}(\q) 
	= \frac{1}{M} \sum_{i=1}^{M} E_{f,i}(\q),
\end{equation}
where $E_{f,i}$ is the number of function evaluations required to find the global optimum for a function $f$, $\q$ are the parameters used for the optimization routine, and $M$ are the total number of simulations which successfully found the global optimum. However, this is not the most useful measure as it does not take into account the rate at which the algorithm successfully finds the global optimum. Indeed, defining $r_f(\q)$ as the rate at which a correct solution is found (to within a specified tolerance) for a set of the method parameters $\q$. Then, it may turn out that $\bar E_f(\q_1) < \bar E_f(\q_2)$, but $r_f(\q_1) \ll r_f(\q_2)$ for some certain $\q_1$ and $\q_2$. It then may happen that we would require many more runs for $\q_1$ so that $\q_2$ was actually the better choice. Therefore, we make a refinement to Eq. \eqref{eq:E_i}, and obtain the optimal parameters $\qopt$ for each optimization method by solving the following auxiliary optimization problem:
%
%
\begin{subequations} \label{eq:aux_opt}
\begin{align}
	\qopt & = \arg \, \min_{\q} \{ F_{f,\alpha}(\q) \}, \label{eq:aux_opt1}  \\
	F_{f,\alpha}(\q) & = N_{f,\alpha}(\q) \bar E_f(\q) \label{eq:aux_opt2} \\
	N_{f,\alpha}(\q) & = \cfrac{\log (1-\alpha)}{\log [1 - r_f(\q)]}, \label{eq:aux_opt3}
\end{align}
\end{subequations}
where $N_{f,\alpha}$ is the number of iterations needed to be at least $\alpha$ sure that we have found the global solution and ${\bar E}_f(\q)$ was defined by Eq.~\eqref{eq:E_i}. In this work, we used $\alpha = 0.99$. The dimensionality in the function was absorbed into the optimization problem, and an analysis of the problem dimensionality and number of iterations was tested.

We sampled $10^3$ different starting configurations $\x$, where $x_i \in [-10,10]$ for the algorithms and then ran the nested PSO algorithm 10 times for each dimension and each benchmark function. The best parameters were recorded. Once these were obtained, we tested all the algorithms mentioned using the same starting points (when possible) and compared the performance. The algorithms each had a maximum iteration number of $10^3 n^2$  where $n = |\Omega|$ (the exception being PSO, where we used $\max (10^3 n^2, 100S)$ to ensure at least 100 possible iterations of PSO) is the dimension of the original optimization problem~\eqref{eq:opt}, and would exit out of the loop with a tolerance of $10^{-3} n^2$.

\subsection{Sphere function}
	
\begin{figure}[tbh]
    \includegraphics[width=8cm]{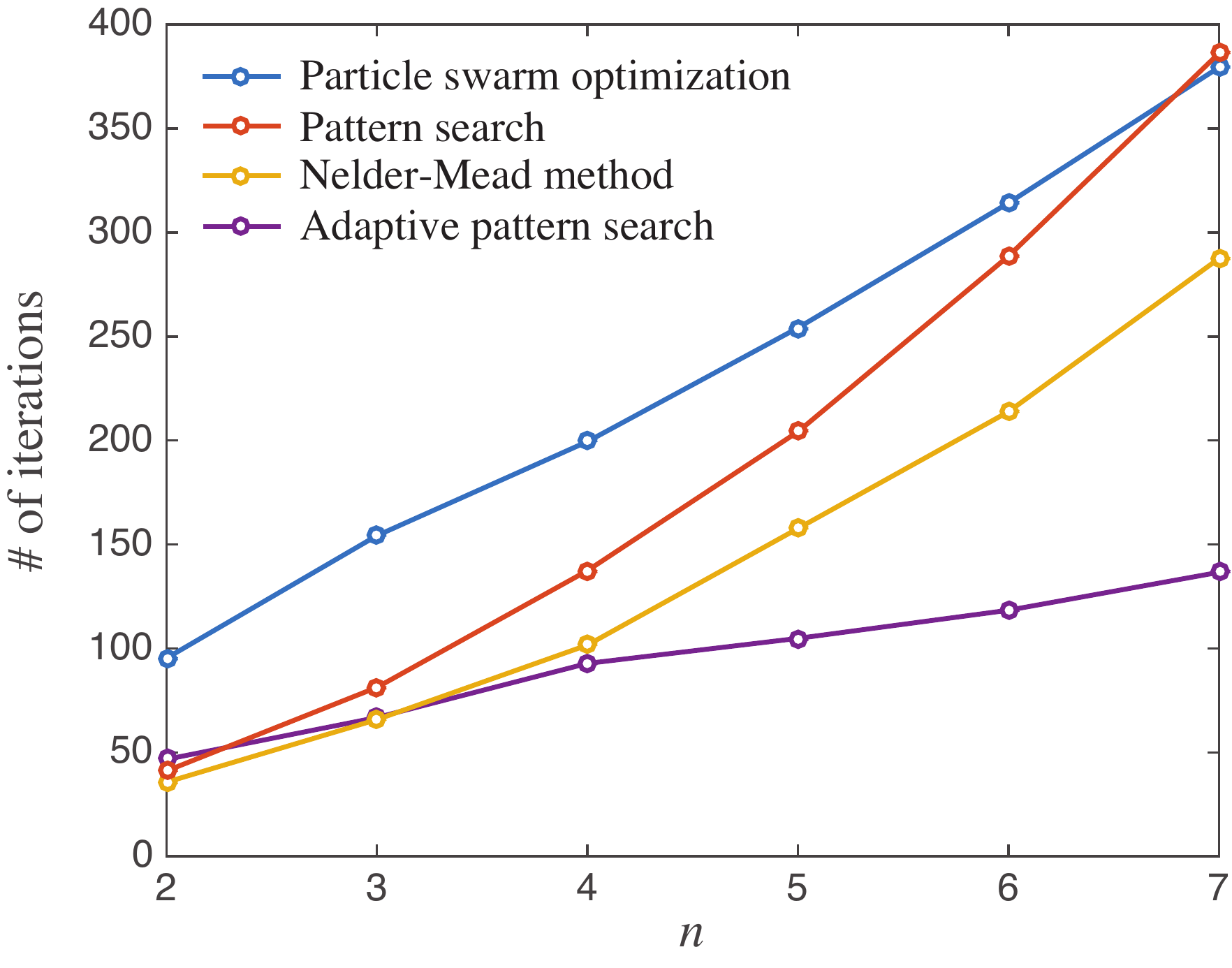}
    \vspace{-3mm}
    \caption{A comparison of  each methods efficiency as a function of dimension for the sphere function. 
    }
\label{fig:sphere_function_comparison}
\end{figure}

The sphere function shown in Fig.~\subref{fig:sphere_function} is defined by
\begin{equation}
	f(\x) = \sum_{i=1}^n x_i^2.
\end{equation}
Tables~\ref{tab:opt_PSO_sphere} and \ref{tab:opt_adaptive_sphere} show a tabulated view of the effectiveness of the chosen methods to find the extremum of this function in the given number of iterations. The function is very simple and the coordinates are uncorrelated. Thus, a wide range of parameters actually turned out to be similarly effective. A comparison of the method's performances is presented in Fig.~\ref{fig:sphere_function_comparison} by using 100 random initial starting configurations and employing Eq.~\eqref{eq:aux_opt}.

\begin{table}[tbh]
\centering\normalsize
\begin{tabularx}{0.9\columnwidth}{@{}l *6{R}@{}}
\toprule
$n$ & $\{ S$ & $\omega$ & $\phi_\mathrm{p}$ & $\phi_\mathrm{g} \}$ & $F_{f,\alpha}$ \vspace{1.5pt}\\
\hline
2 & 5 & 0.22 & 0.93 & 1.93 & 84.7 \\
3 & 5 & 0.36 & 1.35 & 1.68 & 131.3 \\
4 & 8 & 0.23 & 0.80 & 1.96 & 183.7 \\
5 & 10 & 0.18 & 0.99 & 1.96 & 235.5 \\
6 & 9 & 0.36 & 1.55 & 1.55 & 290.5 \\
7 & 12 & 0.27 & 1.18 & 1.75 & 348.4 \\
\bottomrule
\end{tabularx}
\caption{Optimal PSO parameters $\qopt = \{ S$, $\omega$, $\phi_\mathrm{p}$, $\phi_\mathrm{g} \}$ and $F_{f,\alpha}(\qopt)$ for the sphere function.}
\label{tab:opt_PSO_sphere}
\end{table}

\begin{table}[tbh]
\centering\normalsize
\begin{tabularx}{0.9\columnwidth}{@{}l *6{R}@{}}
\toprule
$n$ & $\{ \mu$ & $\sigma$ & $k_\mathrm{s}$ & $k_\mathrm{u} \}$ & $F_{f,\alpha}$ \vspace{1.5pt}\\
\hline
2 & 1 & 0.21 & 1.00 & 0.24 & 44.3 \\
3 & 2 & 0.22 & 1.00 & 0.33 & 64.0 \\
4 & 4 & 0.12 & 1.00 & 0.29 & 82.0 \\
5 & 2 & 0.36 & 1.00 & 0.29 & 99.5 \\
6 & 3 & 0.21 & 1.00 & 0.33 & 114.9 \\
7 & 13 & 0.25 & 1.00 & 0.33 & 128.7 \\
\bottomrule
\end{tabularx}
\caption{Optimal adaptive pattern search parameters $\qopt = \{ \mu$, $\sigma$, $k_\mathrm{s}$, $k_\mathrm{u} \}$ and $F_{f,\alpha}(\qopt)$ for the sphere function.}
\label{tab:opt_adaptive_sphere}
\end{table}

\subsection{Rosenbrock function}

The Rosenbrock function shown in Fig.~\subref{fig:Rosenbrock_function} is a standard test for optimization methods and is given by
\begin{equation}
	f(\x) = \sum_{i=1}^{n-1} \bigl[ 100(x_{i+1}-x_i^2)^2 + (1-x_i)^2 \bigr].
\end{equation}
Optimization of this function demonstrates the utility of coordinate independent local search methods (like Nelder-Mead or adaptive pattern search). This is due to the fact that the minimum is contained inside a parabolic valley which requires constant shrinking of the step size for pattern search to make progress. Consider the 2D case
\begin{equation*}
	f(x,y) = 100(y-x^2)^2 + (1-x)^2.
\end{equation*}
A change of variables to the $(u,v)$ plane given by $u = 1 - x$ and $v = y - x^2$ leads to elliptical level sets $g(u,v) = u^2 + 100 v^2$, which is much more favorable to coordinate-dependent methods. In practice it is usually difficult or impossible to find the appropriate transform converting to elliptical level sets. Nevertheless, this gives a useful test of the morphology of the surface of~$\Jc$ by comparing iterations between coordinate-dependent and coordinate-independent methods.

\begin{figure}[tbh]
	\includegraphics[width=8cm]{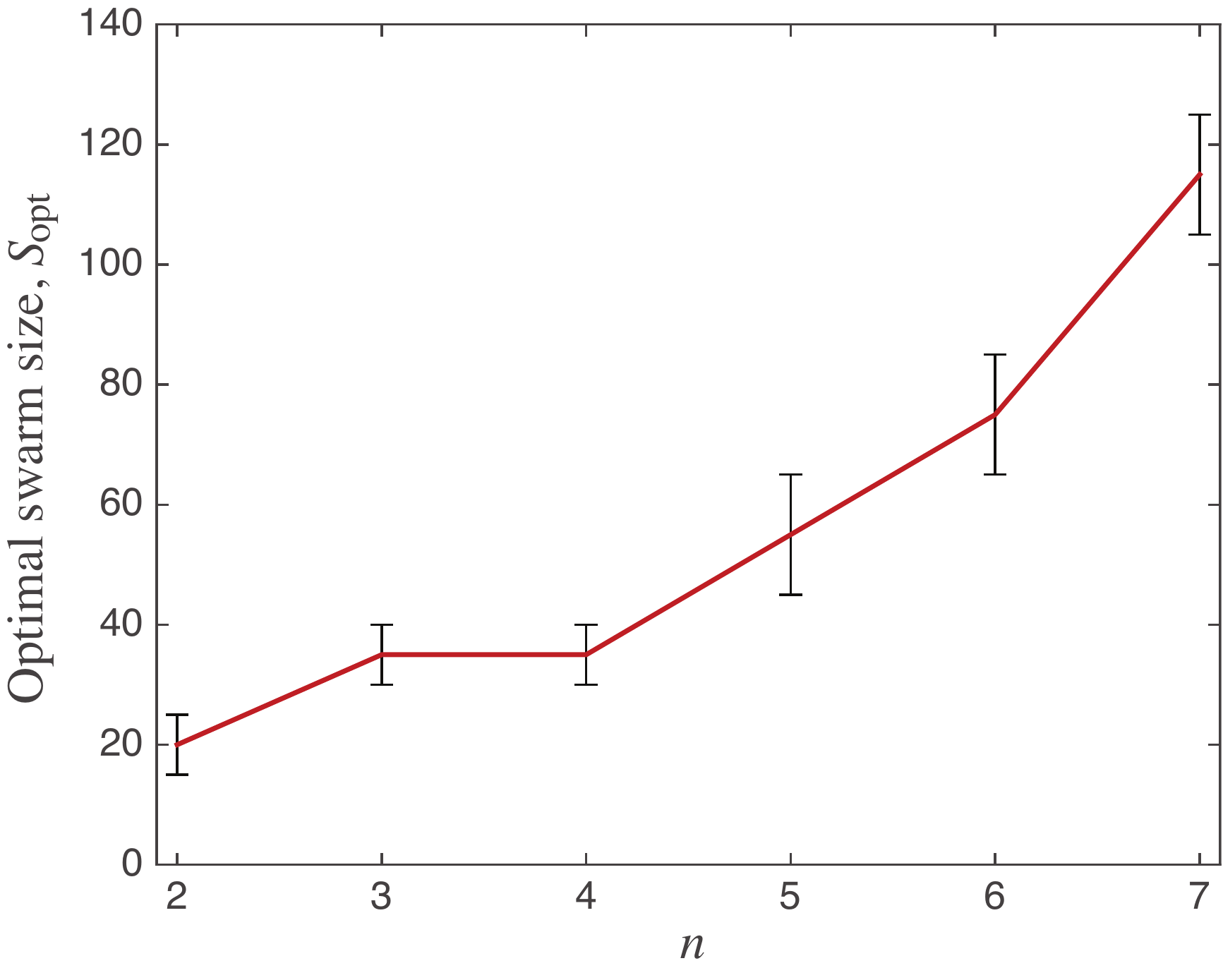}
    \vspace{-3mm}
	\caption{The optimal swarm size, $S$, of the particle swam optimization on the Rosenbrock function as a function of dimension, $n$ while holding the other parameters fixed $\{ \omega$, $\phi_\mathrm{p}$, $\phi_\mathrm{g} \} = \{0.3$, $-0.2$, $1.7 \}$.}
\label{fig:Rosenbrock_swarm_size}
\end{figure}

Tables~\ref{tab:opt_PSO_Rosenbrock} and \ref{tab:opt_adaptive_Rosenbrock} show the effectiveness of the PSO and adaptive pattern search methods to minimize the Rosenbrock function in a given number of iterations. As we can see from the table, the PSO method is much slower than Nelder-Mead and adaptive pattern search. The optimization of parameters for PSO have revealed that the dimensionality and swarm size are (perhaps not surprisingly) correlated. As the dimensionality increases, the optimal swarm size (holding other parameters fixed) increases. To verify this, we sampled 100 random starting points for swarm sizes between 10--200; the results are shown in Fig.~\ref{fig:Rosenbrock_swarm_size}.

\begin{table}[tbh]
\centering\normalsize
\begin{tabularx}{1.\columnwidth}{@{}l *8{R}@{}}
\toprule
$n$ & $\{ S$ & $\omega$ & $\phi_\mathrm{p}$ & $\phi_\mathrm{g} \}$ & $F_{f,\alpha}$ & $r_f$ & $N_{f,\alpha}$ \vspace{1.5pt}\\
\hline
2 & 28 & 0.25 & $-0.21$ & 1.58 & 823 & 0.99 & 1 \\
3 & 34 & 0.29 & $-0.15$ & 1.73 & 2942 & 0.95 & 2 \\
4 & 40 & 0.35 & $-0.17$ & 1.59 & 9352 & 0.73 & 4 \\
5 & 49 & 0.28 & $-0.21$ & 1.67 & 15650 & 0.61 & 5 \\
6 & 61 & 0.30 & $-0.20$ & 1.70 & 22339 & 0.56 & 6 \\
7 & 73 & 0.27 & $-0.19$ & 1.68 & 39277 & 0.53 & 7 \\
\bottomrule
\end{tabularx}
\caption{Optimal PSO parameters $\qopt = \{ S$, $\omega$, $\phi_\mathrm{p}$, $\phi_\mathrm{g} \}$, $F_{f,\alpha}(\qopt)$, $r_f(\qopt)$, and $N_{f,\alpha}(\qopt)$ for the Rosenbrock function.}
\label{tab:opt_PSO_Rosenbrock}
\end{table}

\begin{table}[tbh]
\centering\normalsize
\begin{tabularx}{1.\columnwidth}{@{}l *8{R}@{}}
\toprule
$n$ & $\{ \mu$ & $\sigma$ & $k_\mathrm{s}$ & $k_\mathrm{u} \}$ & $F_{f,\alpha}$ & $r_f$ & $N_{f,\alpha}$ \vspace{1.5pt}\\
\hline
2 & 2 & 0.28 & 1.67 & 0.43 & 227 & 1.00 & 1 \\
3 & 3 & 0.15 & 1.48 & 0.46 & 392 & 1.00 & 1 \\
4 & 4 & 0.22 & 1.81 & 0.38 & 781 & 0.94 & 2 \\
5 & 5 & 0.14 & 1.71 & 0.45 & 1384 & 0.92 & 2 \\
6 & 6 & 0.37 & 1.96 & 0.36 & 1775 & 0.93 & 2 \\
7 & 7 & 0.10 & 2.77 & 0.34 & 2993 & 0.92 & 2 \\
\bottomrule
\end{tabularx}
\caption{Optimal adaptive pattern search parameters $\qopt = \{ \mu$, $\sigma$, $k_\mathrm{s}$, $k_\mathrm{u} \}$, $F_{f,\alpha}(\qopt)$, $r_f(\qopt)$, and $N_{f,\alpha}(\qopt)$ for the Rosenbrock function.}
\label{tab:opt_adaptive_Rosenbrock}
\end{table}

\begin{figure}[tbh]
    \includegraphics[width=8cm]{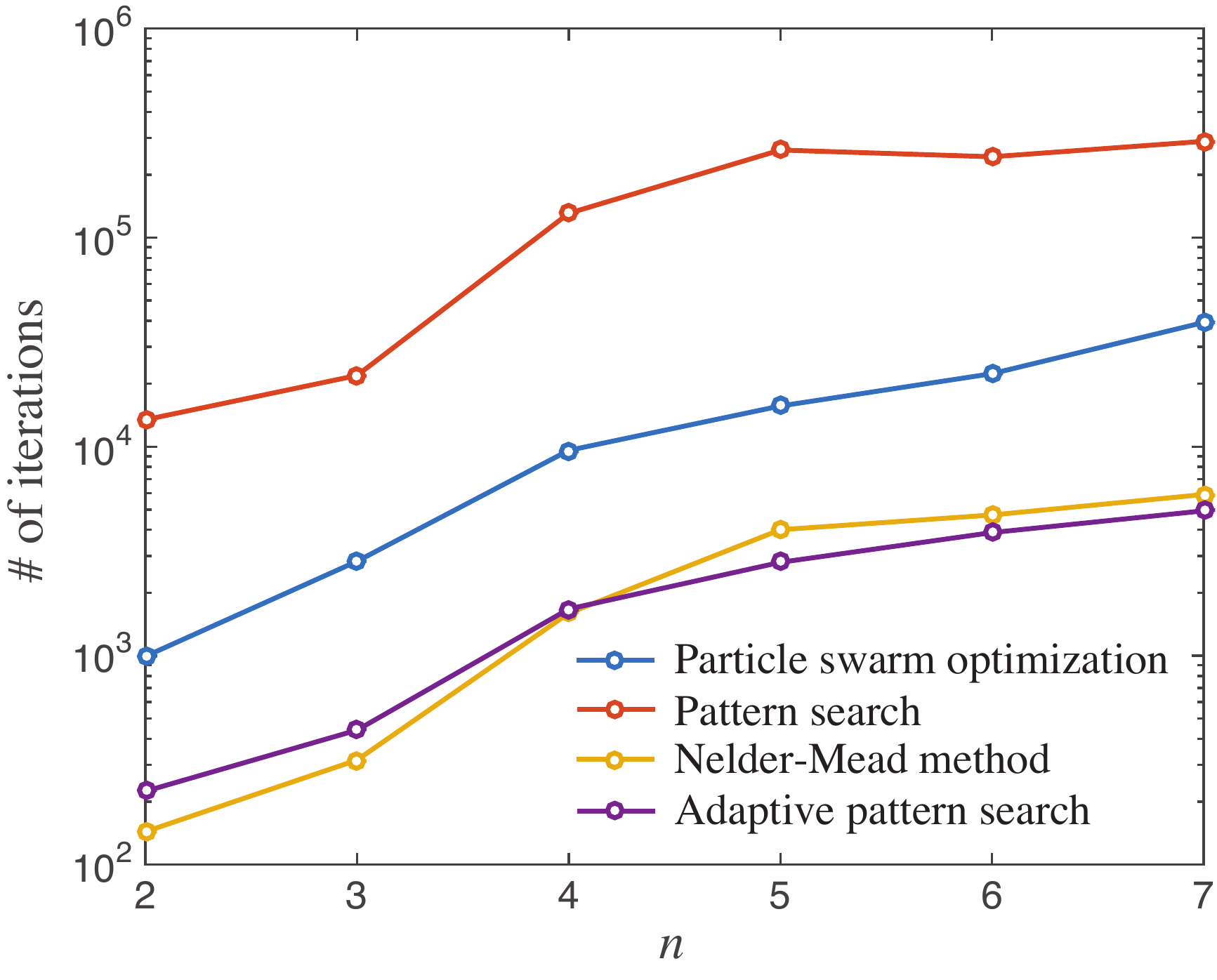}
    \vspace{-3mm}
    \caption{A comparison of each methods efficiency as a function of dimension for the Rosenbrock function. 
    }
\label{fig:Rosenbrock_function_comparison}
\end{figure}

\subsection{Rastrigin function}

\begin{figure}[tbh]
    \includegraphics[width=8cm]{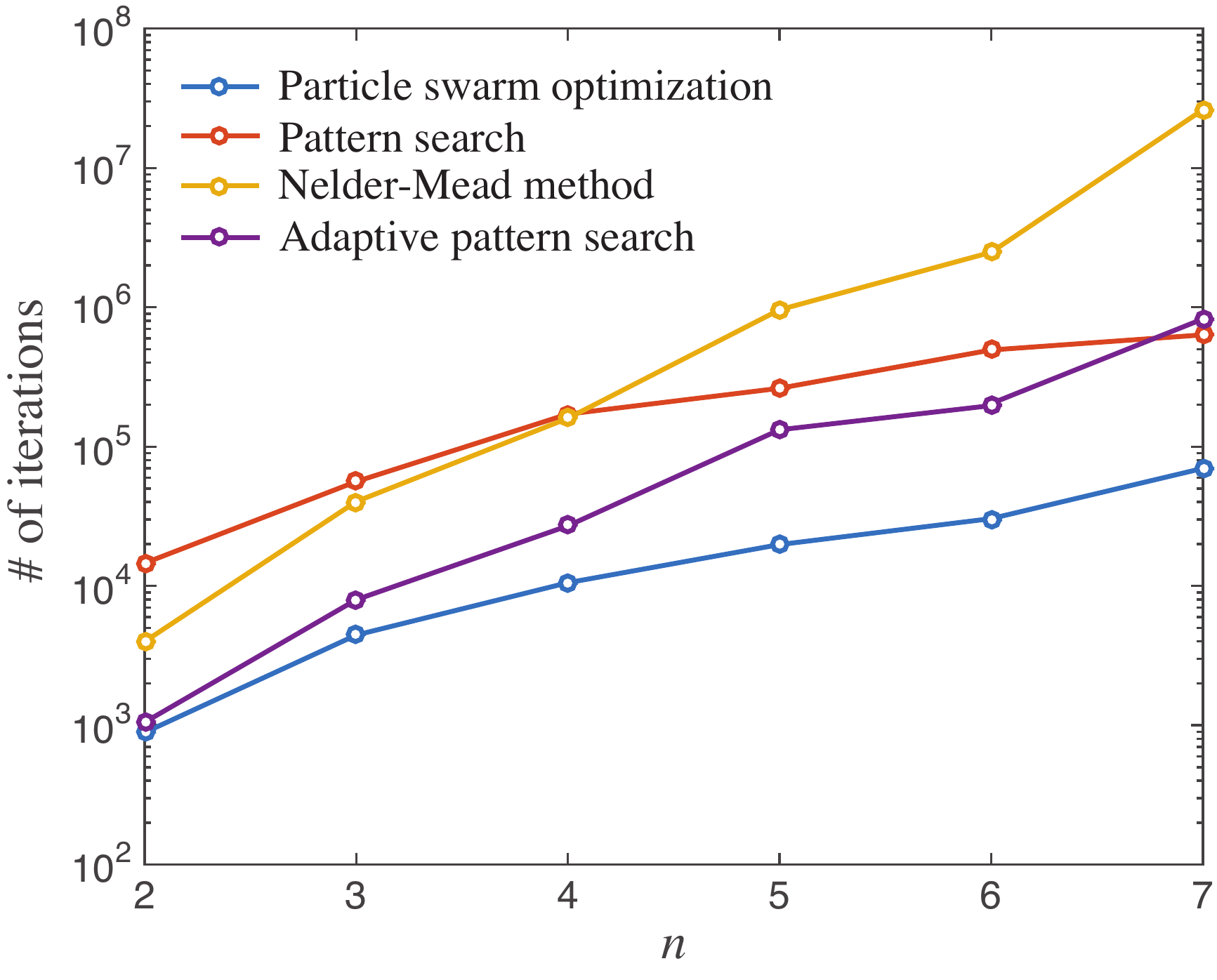}
    \vspace{-3mm}
    \caption{A comparison of each methods efficiency as a function of dimension for the Rastrigin function. 
    }
\label{fig:Rastrigin_function_comparison}
\end{figure}

\begin{figure}[tbh]
	\includegraphics[width=8cm]{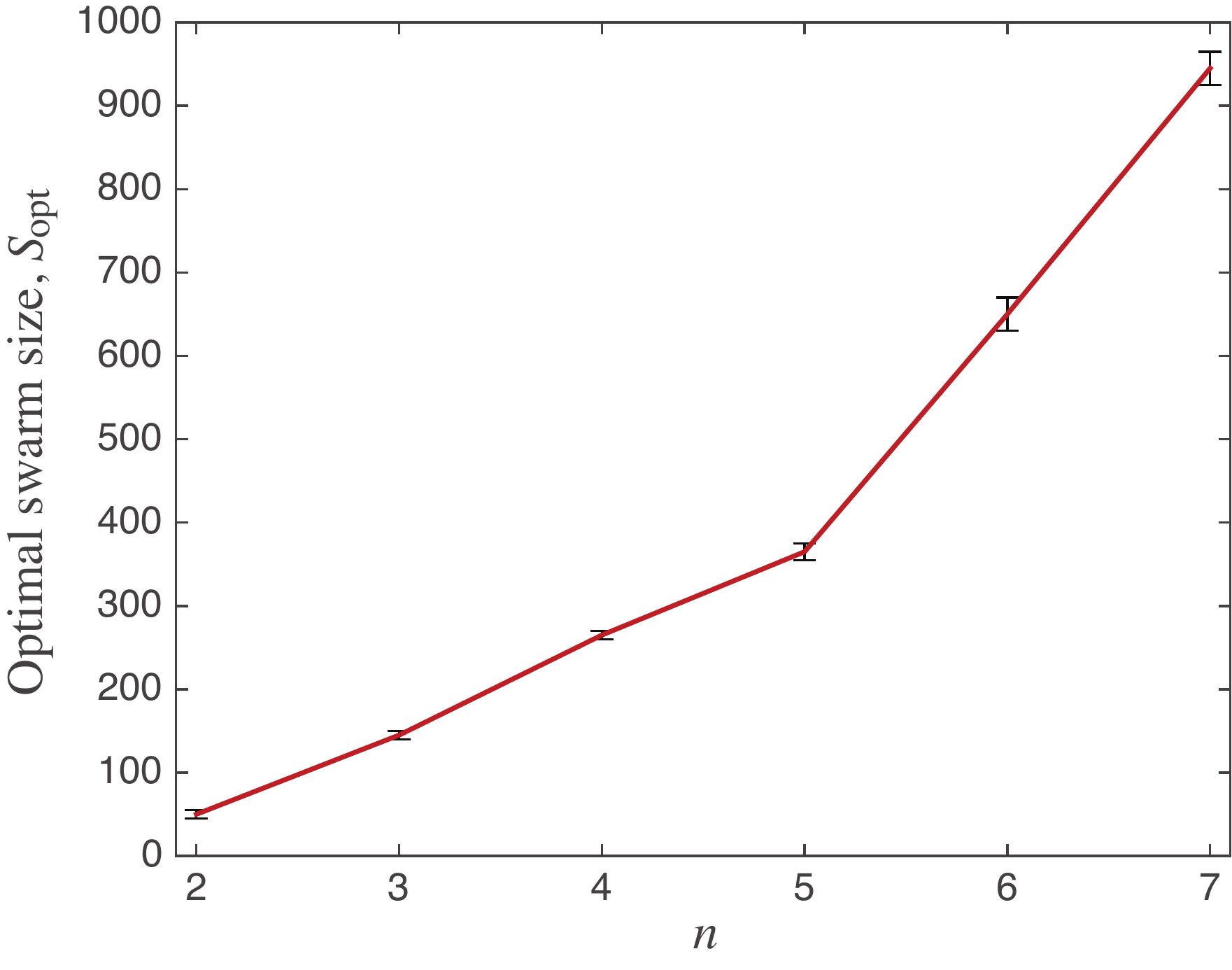}
    \vspace{-3mm}
	\caption{The optimal swarm size, $S$, of the particle swam optimization on the Rastrigin function as a function of dimension, $n$ while holding the other parameters fixed $(\omega$, $\phi_\mathrm{p}$, $\phi_\mathrm{g}) = (0.4$, $2.0$, $1.0)$.}
\label{fig:Rastrigin_swarm_size}
\end{figure}

The third test function we consider here, is the  Rastrigin function [see Fig.~\subref{fig:Rastrigin_function}], which is the sum of a sphere function and periodic oscillations, having multiple minima. This type of objective function is similar to the objective function of superconductors having periodic pinscapes. It is defined by 
\begin{equation}
	f(\x) = 10n + \sum_{i=1}^n \bigl[ x_i^2 - 10 \cos(2 \pi x_i) \bigr].
\end{equation}
This is where PSO performs particularly well compared to the local search methods. Table~\ref{tab:opt_PSO_Rastrigin} shows its effectiveness to solve the Rastrigin function optimization problem in the given number of iterations. The local methods do a comparatively poor job as can be seen in Table~\ref{tab:opt_adaptive_Rastrigin} and Fig.~\ref{fig:Rastrigin_function_comparison}. The optimal swarm size also grows rapidly with increasing dimensionality and we sampled 100 random starting points for swarm sizes between 50 and 1,000; the results are shown in Fig.~\ref{fig:Rastrigin_swarm_size}.

\begin{table}[tbh]
\centering\normalsize
\begin{tabularx}{1.\columnwidth}{@{}l *8{R}@{}}
\toprule
$n$ & $\{ S$ & $\omega$ & $\phi_\mathrm{p}$ & $\phi_\mathrm{g} \}$ & $F_{f,\alpha}$ & $r_f$ & $N_{f,\alpha}$ \vspace{1.5pt}\\
\hline
2 & 50 & 0.25 & 2.00 & 1.00 & 934 & 1.00 & 1 \\
3 & 145 & 0.58 & 2.08 & 0.89 & 4514 & 1.00 & 1 \\
4 & 265 & 0.58 & 2.21 & 0.75 & 10953 & 1.00 & 1 \\
5 & 365 & 0.48 & 2.22 & 0.77 & 20410 & 1.00 & 1 \\
6 & 650 & 0.5 & 2.25 & 0.67 & 43939 & 0.998 & 1 \\
7 & 945 & 0.51 & 2.29 & 0.65 & 81779 & 0.993 & 1 \\
\bottomrule
\end{tabularx}
\caption{Optimal PSO parameters $\qopt = \{ S$, $\omega$, $\phi_\mathrm{p}$, $\phi_\mathrm{g} \}$, $F_{f,\alpha}(\qopt)$, $r_f(\qopt)$, and $N_{f,\alpha}(\qopt)$ for Rastrigin function.}
\label{tab:opt_PSO_Rastrigin}
\end{table}

\begin{table}[tbh]
\centering\normalsize
\begin{tabularx}{1.\columnwidth}{@{}l *8{R}@{}}
\toprule
$n$ & $\{ \mu$ & $\sigma$ & $k_\mathrm{s}$ & $k_\mathrm{u} \}$ & $F_{f,\alpha}$ & $r_f$ & $N_{f,\alpha}$ \vspace{1.5pt}\\
\hline
2 & 3 & 0.25 & 1.79 & 0.33 & 1051 & 0.22 & 19 \\
3 & 1 & 0.37 & 1.94 & 0.31 & 7972 & 0.04 & 113 \\
4 & 1 & 0.39 & 1.69 & 0.29 & 27017 & 0.01 & 327 \\
5 & 2 & 0.62 & 1.00 & 0.45 & 132134 & 0.004 & 1149 \\
6 & 6 & 0.3 & 6.59 & 0.29 & 197052 & 0.003 & 1532 \\
7 & 7 & 0.41 & 5.93 & 0.11 & 584564 & 0.001 & 4603 \\
\bottomrule
\end{tabularx}
\caption{Optimal adaptive search parameters $\qopt = \{ S$, $\omega$, $\phi_\mathrm{p}$, $\phi_\mathrm{g} \}$, $F_{f,\alpha}(\qopt)$, $r_f(\qopt)$, and $N_{f,\alpha}(\qopt)$ for Rastrigin function.}
\label{tab:opt_adaptive_Rastrigin}
\end{table}

\section{Model for the critical current in superconductors} \label{sec:GL}

Here we define the objective function for the optimization problem in superconductors mentioned before.
For the description of the vortex dynamics in strong {type-II} superconductors, we use the TDGL equation for the superconducting order parameter $\psi = \psi(\mathbf{r},t)$,
\begin{equation} 
	(\partial_t + i\mu)\psi 
	= \epsilon (\mathbf{r})\psi - |\psi |^2\psi + 
	(\nabla - i\mathbf{A})^2\psi + \zeta(\mathbf{r},t),
	\label{eq:GL} 
\end{equation} 
which is solved numerically. Here $\mu = \mu(\mathbf{r},t)$ is the chemical potential, $\mathbf{A}$ is the vector potential associated with an external magnetic field $\mathbf{B}$ as $\mathbf{B} = \nabla \times \mathbf{A}$, and $\zeta(\mathbf{r},t)$ is the temperature-dependent $\delta$-correlated Langevin term. The unit of length is given by the superconducting coherence length~$\xi$ and the unit of magnetic field is given by the upper critical field $\Hct$. 
See Ref.~\onlinecite{Sadovskyy:2015a} for the details of TDGL model implementation and physical units.
The current density is given by the expression 
\begin{equation}
	\mathbf{J} 
	= \mathrm{Im} \bigl[ \psi^*(\nabla - i\mathbf{A})\psi \bigr]
	- \nabla \mu.
	\label{eq:J} 
\end{equation} 

To determine the critical current value~--- the maximal current, which can flow through the superconductor without dissipation, --- we use a finite-electrical-field criterion. Specifically, we choose a certain small external electric field,  $\Ec = 10^{-4}$, which measures the dissipation and adjust the applied external current, $\Jext$, to reach this electrical-field/dissipation level on average during the simulation. The time-averaged value of external current over a steady state gives the critical current, $\Jc = \langle\Jext\rangle$. 

The critical current in the presence of an external magnetic field is mostly defined by the pattern of non-superconducting defects, including their sizes, shapes, and spatial distribution, which prevents vortices from moving under the influence of the Lorentz force, $\fL = \mathbf{J} \times \mathbf{B}$.

The pinscape is characterized by a set of parameters~$\x$, which corresponds to the objective function 
\begin{equation}
	f(\x) = - \Jc(\x) \label{eq:objfunc}
\end{equation}
used in optimization problem \eqref{eq:opt}. Each element $\x$ of the parameter space $\Omega$ describes the pinscape in the superconductor, e.g. the shape of each defect and their spatial distribution. The optimal configuration of the defects $\xopt$ corresponds to the minimization of the objective function, $\fopt = f(\xopt)$.

Knowledge of the shape and behavior of the function $\Jc(\x)$ is not known \textit{a priori}. 
In addition if we consider, for example, a random placement of defects in the domain, each realization can yield slightly different values for~$\Jc$~--- for the same $\x$ (if it does not describe the positions of all defects explicitly), such that averaging is required. In that case one can expect, as the number of random simulations tends to infinity,~$\Jc$ approaches a ``true'' value due to self-averaging. For a finite number of trials it leads, however, to a noisy~$\Jc$ surface. This can create difficulties for local methods to converge to the global solution. A modification of the local methods to multi-level starting points, helps to overcome the noise in these types of problems.

\begin{figure*}
	\hspace{-0.15cm}
	\subfloat{\includegraphics[width=17.8cm]{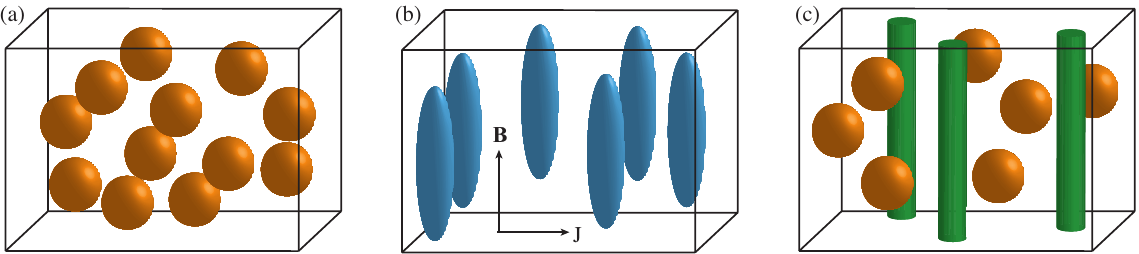} \label{fig:pattern_spherical}}
	\subfloat{\label{fig:pattern_spheroidal}}
	\subfloat{\label{fig:pattern_spherical_columnar}}
	\vspace{-0.25cm}
	\caption{
		Pinscapes for optimization. 
		(a)~Randomly placed spherical inclusions for problem~\eqref{eq:opt2}.
		(b)~Spheroidal inclusions for problem~\eqref{eq:opt3}.
		(c)~Mixture of spherical and columnar inclusions for problem~\eqref{eq:opt4}.
	}
\label{fig:patterns}
\end{figure*}

\begin{figure*}
	\hspace{-0.15cm}
	\begin{tabular}{cc}
	\includegraphics[width=8.8cm]{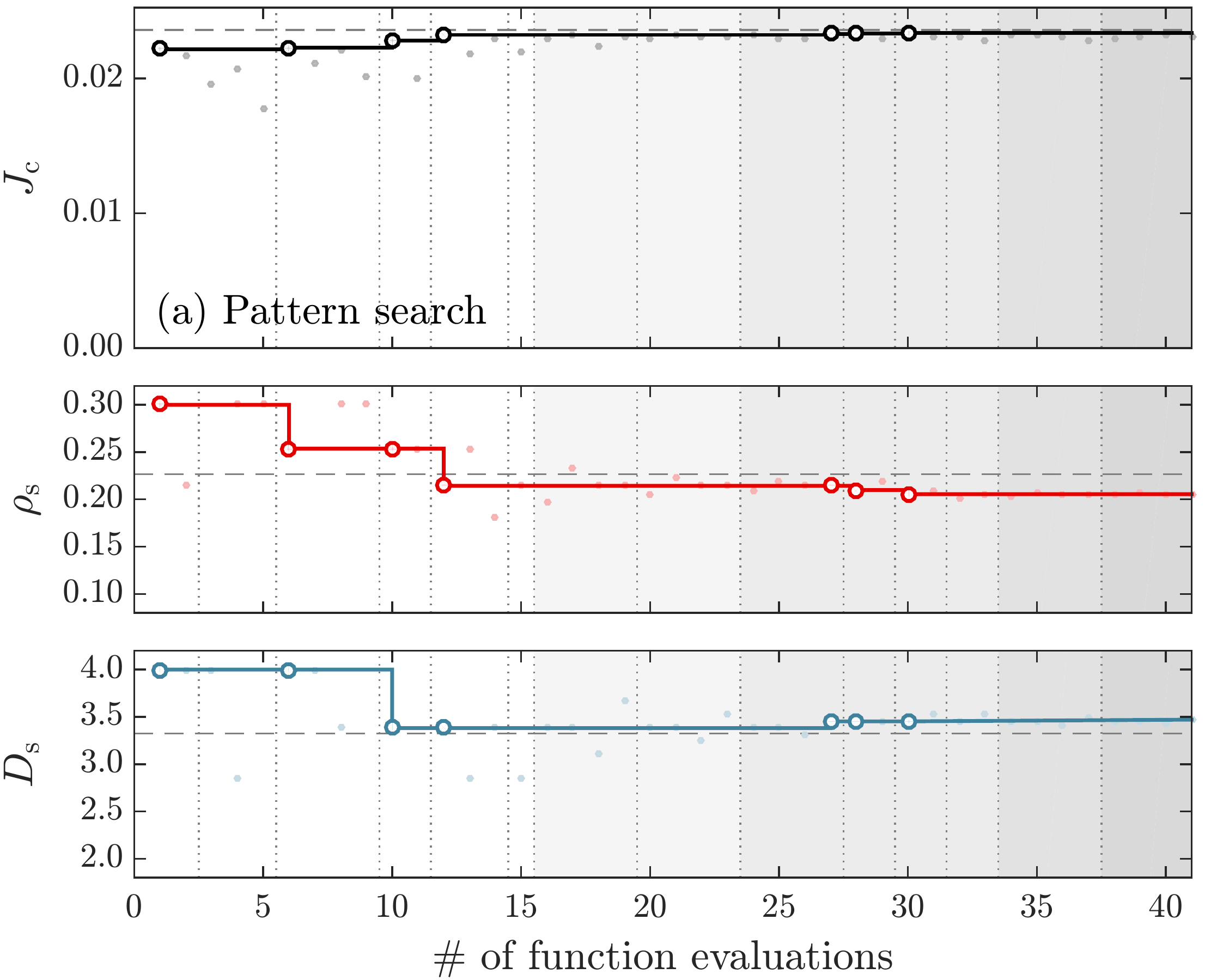} \vspace{0.2cm} &
	\includegraphics[width=8.8cm]{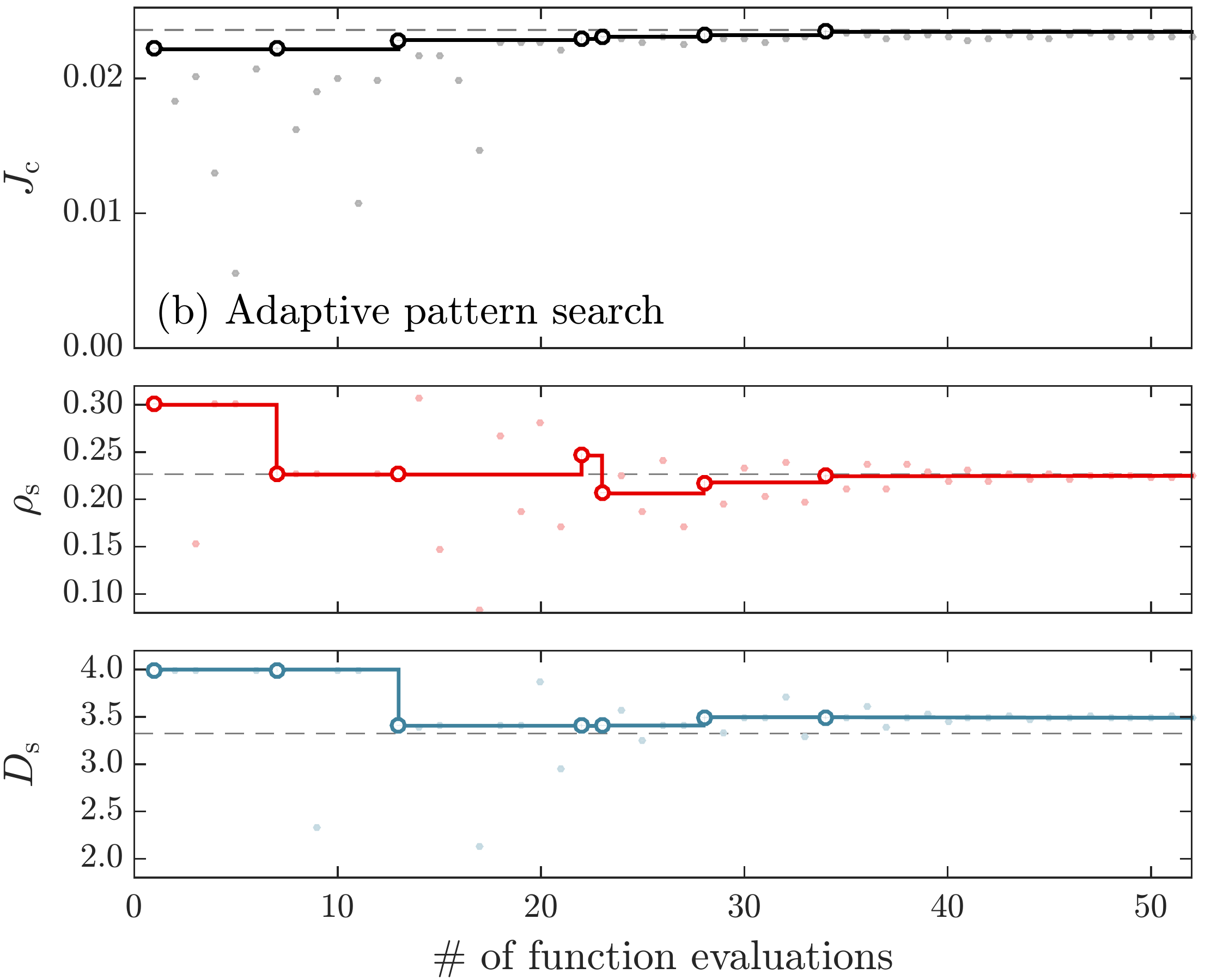} \vspace{0.2cm} \\
	\includegraphics[width=8.8cm]{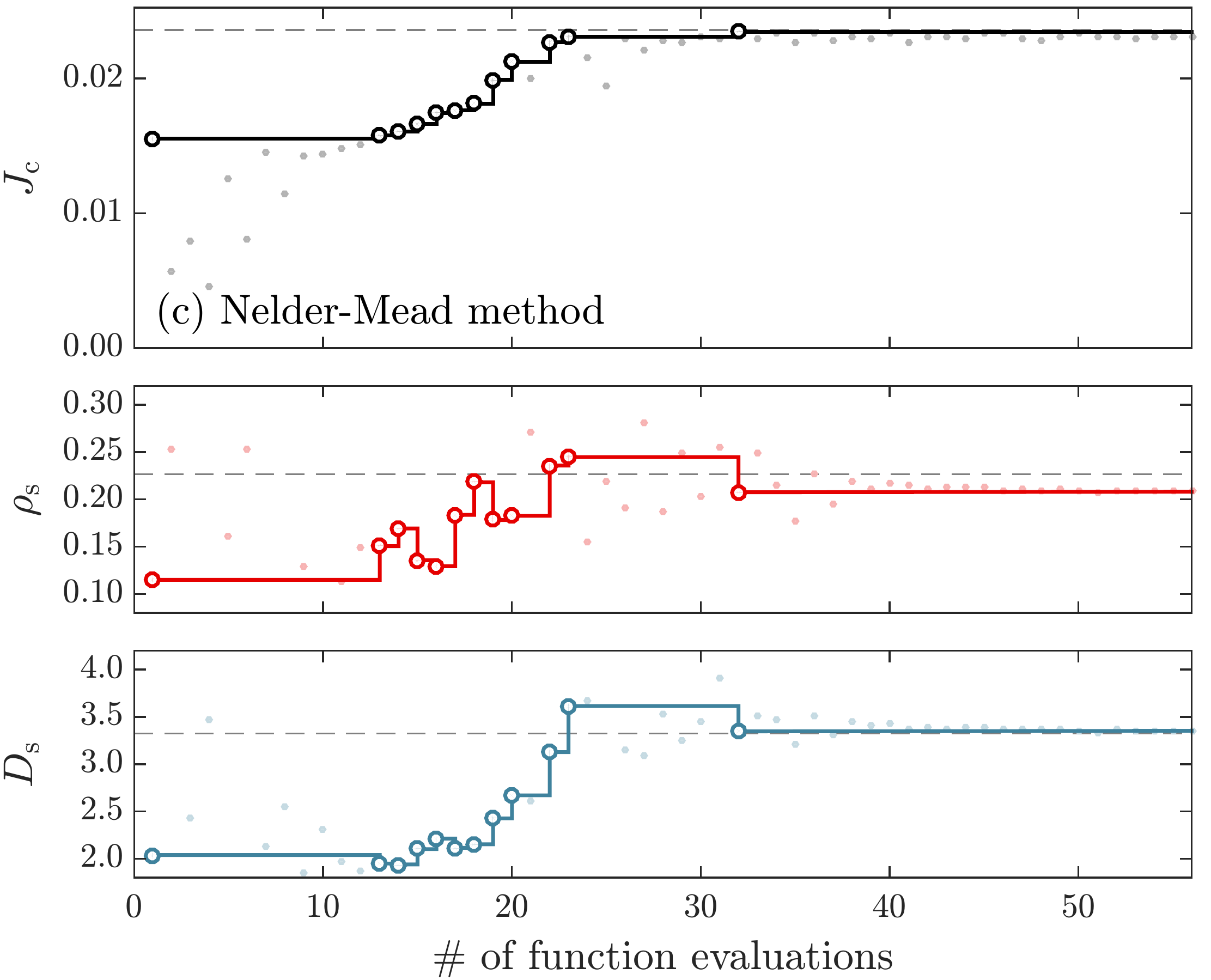} &
	\includegraphics[width=8.8cm]{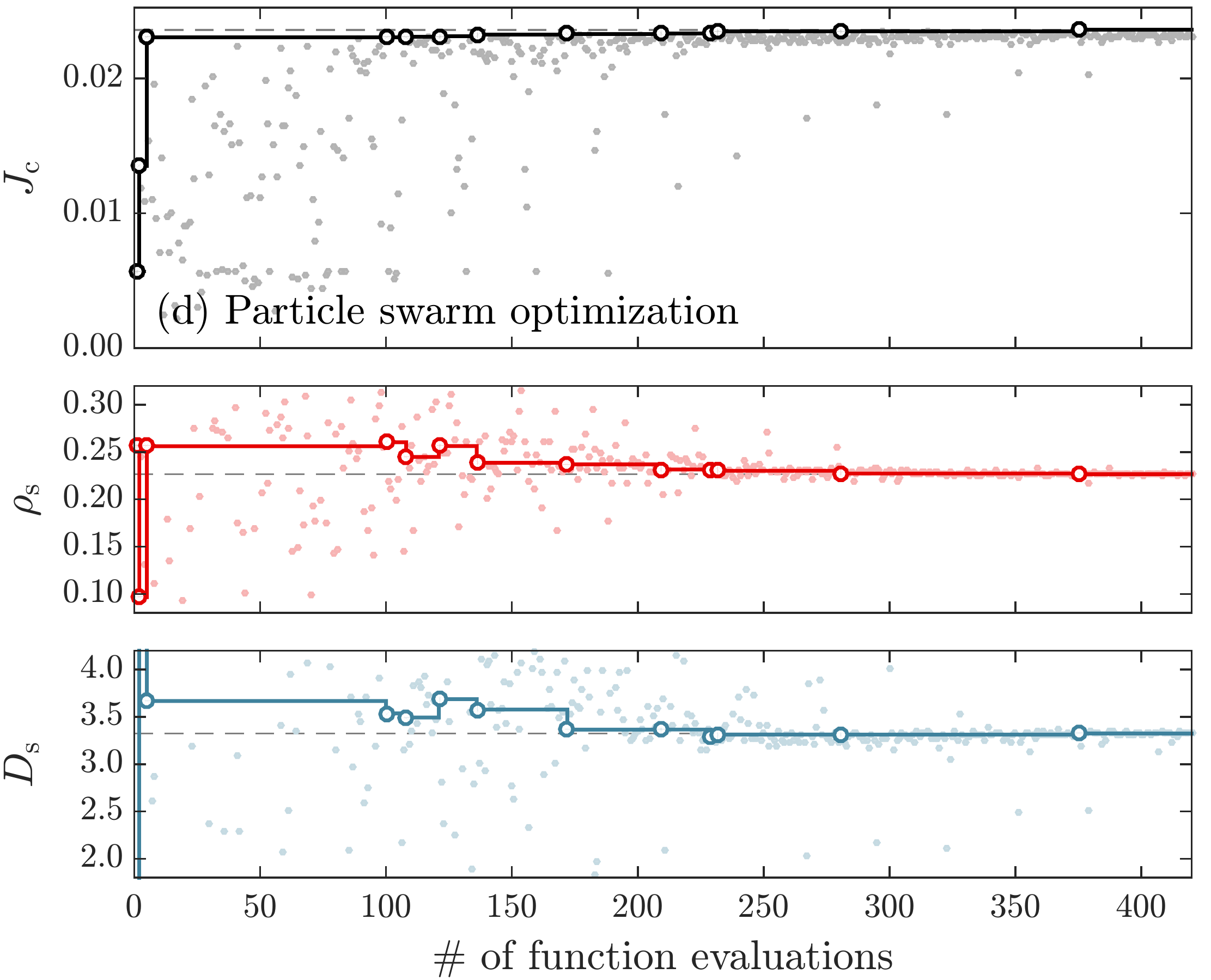}
	\end{tabular}
	\vspace{-0.2cm}
	\caption{
		Optimization procedure for the two-parameter (2D) problem~\eqref{eq:opt2}
		with monodisperse spherical defects characterized by the
		volume fraction $\fs$ occupied by them and their diameter $\Ds$.
		Critical current $\Jc$ and optimization parameters are shown as 
		a function of number of objective function evaluations for
		(a)~pattern search, 
		(b)~adaptive pattern search, 
		(c)~Nelder-Mead method, and 
		(d)~particle swarm optimization.
		All methods converged to the same optimum marked by horizontal dashed line. 
		Vertical dotted lines in panel~(a) separate different iterations; 
		darker background color corresponds to smaller step size.
		PSO exit criterion causes over 300 additional evaluations for marginal improvement.
	}
\label{fig:d2}
\end{figure*}

\begin{figure*}
	\hspace{-0.15cm}
	\begin{tabular}{cc}
	\includegraphics[width=8.8cm]{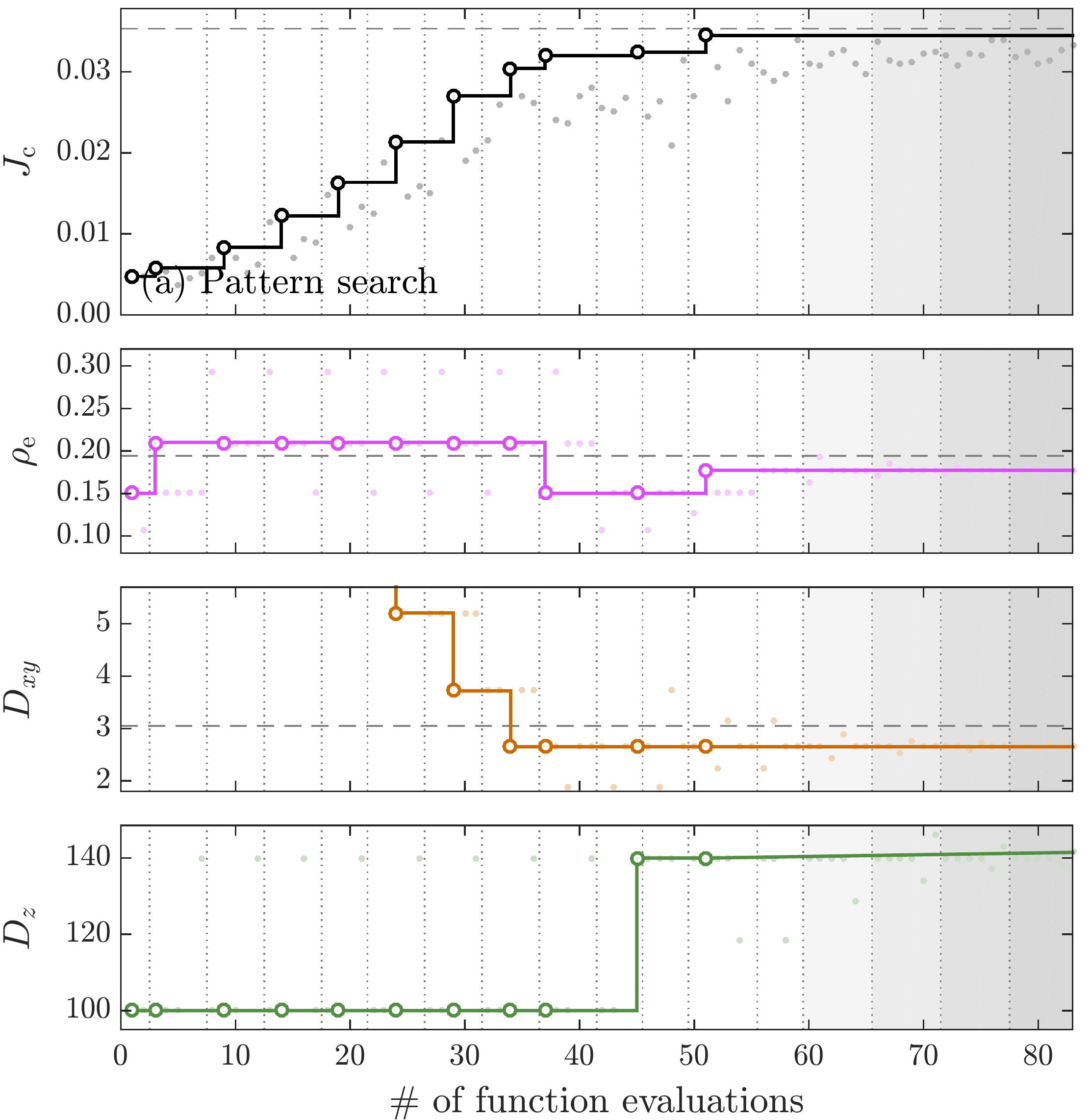} \vspace{0.2cm} &
	\includegraphics[width=8.8cm]{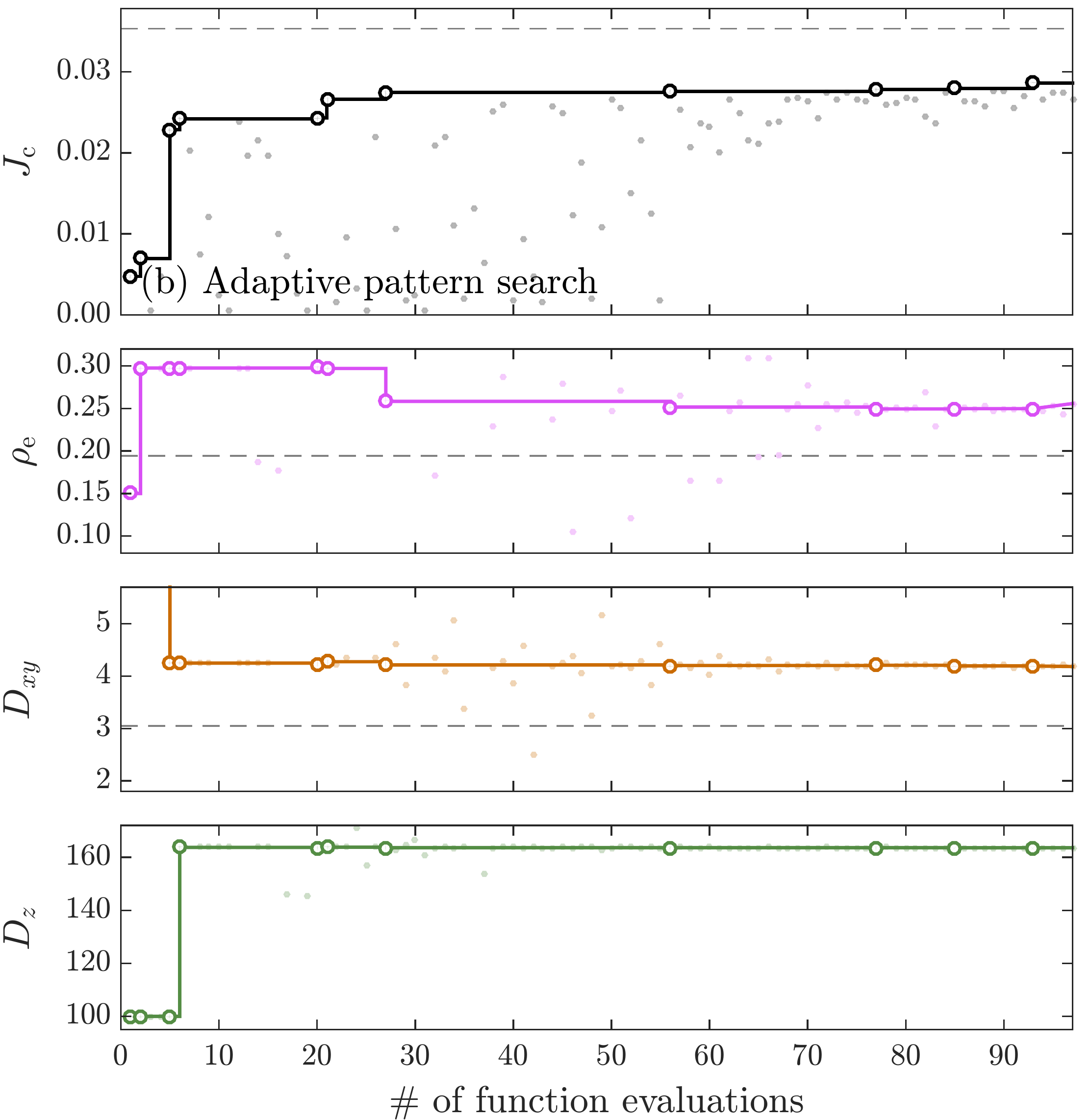} \vspace{0.2cm} \\
	\includegraphics[width=8.8cm]{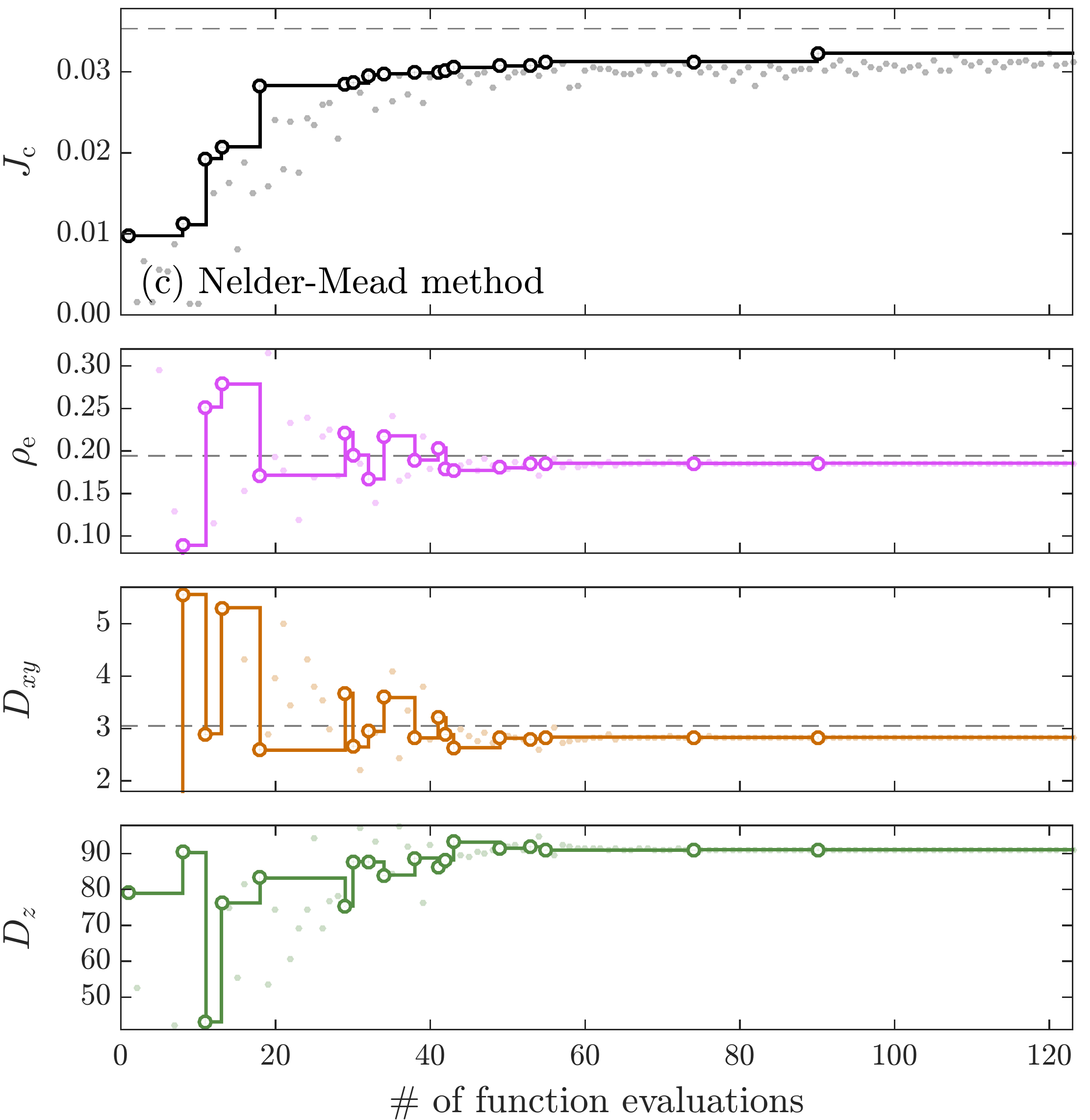} &
	\includegraphics[width=8.8cm]{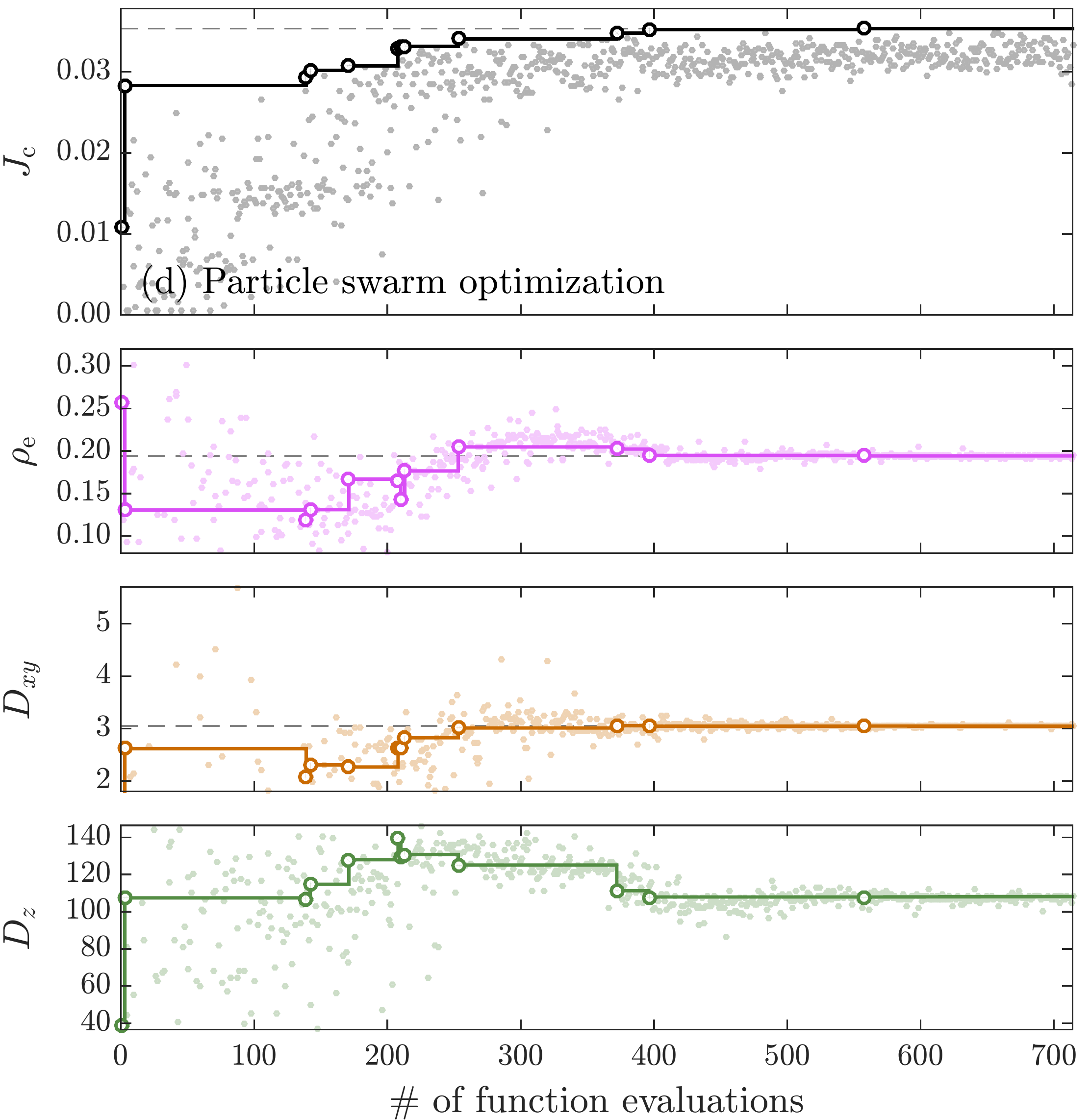}
	\end{tabular}
	\vspace{-0.2cm}
	\caption{
		Optimization procedure for the three-parameter (3D) problem~\eqref{eq:opt3}
		with spheroidal defects characterized by the 
		occupied volume fraction $\fe$ as well as defect diameter in $\Dxy$ in $xy$ plane along applied current and diameter in $\Dz$ in $z$ direction along applied magnetic field.
		Critical current $\Jc$ and optimization parameters are shown as 
		a function of number of objective function evaluations for
		(a)~pattern search, 
		(b)~adaptive pattern search, 
		(c)~Nelder-Mead method, and 
		(d)~particle swarm optimization.
		Optimal diameter in $z$ direction is larger then system size $L_z = 64$ for all methods,
		which indicates that  in the bulk sample optimal $z$-diameter $\Dz$ is infinite and spheroidal defects transform into columnar defects along $z$-axis.
	}
\label{fig:d3}
\end{figure*}

\begin{figure*}
	\hspace{-0.15cm}
	\begin{tabular}{cc}
	\includegraphics[width=8.8cm]{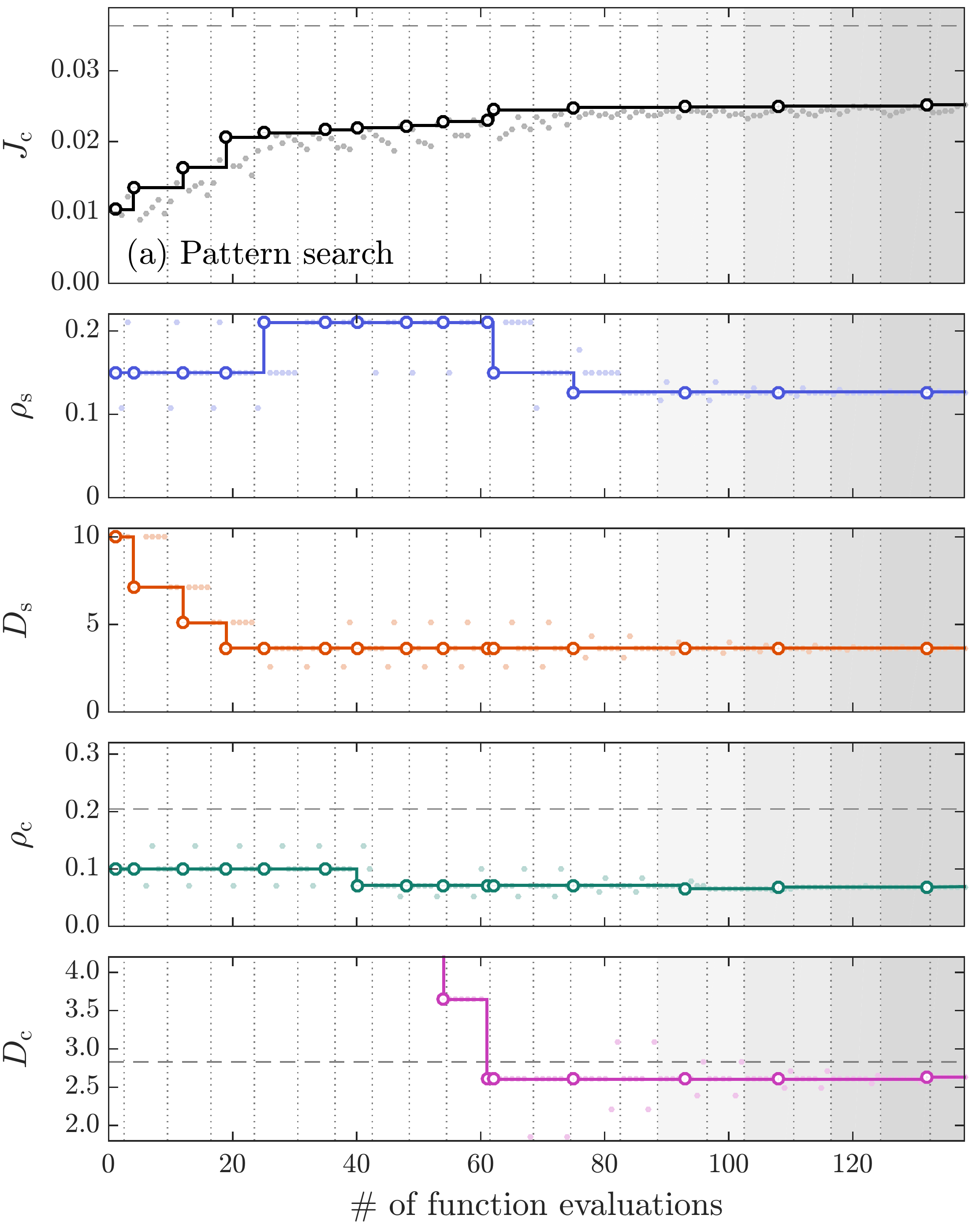} \vspace{0.2cm} &
	\includegraphics[width=8.8cm]{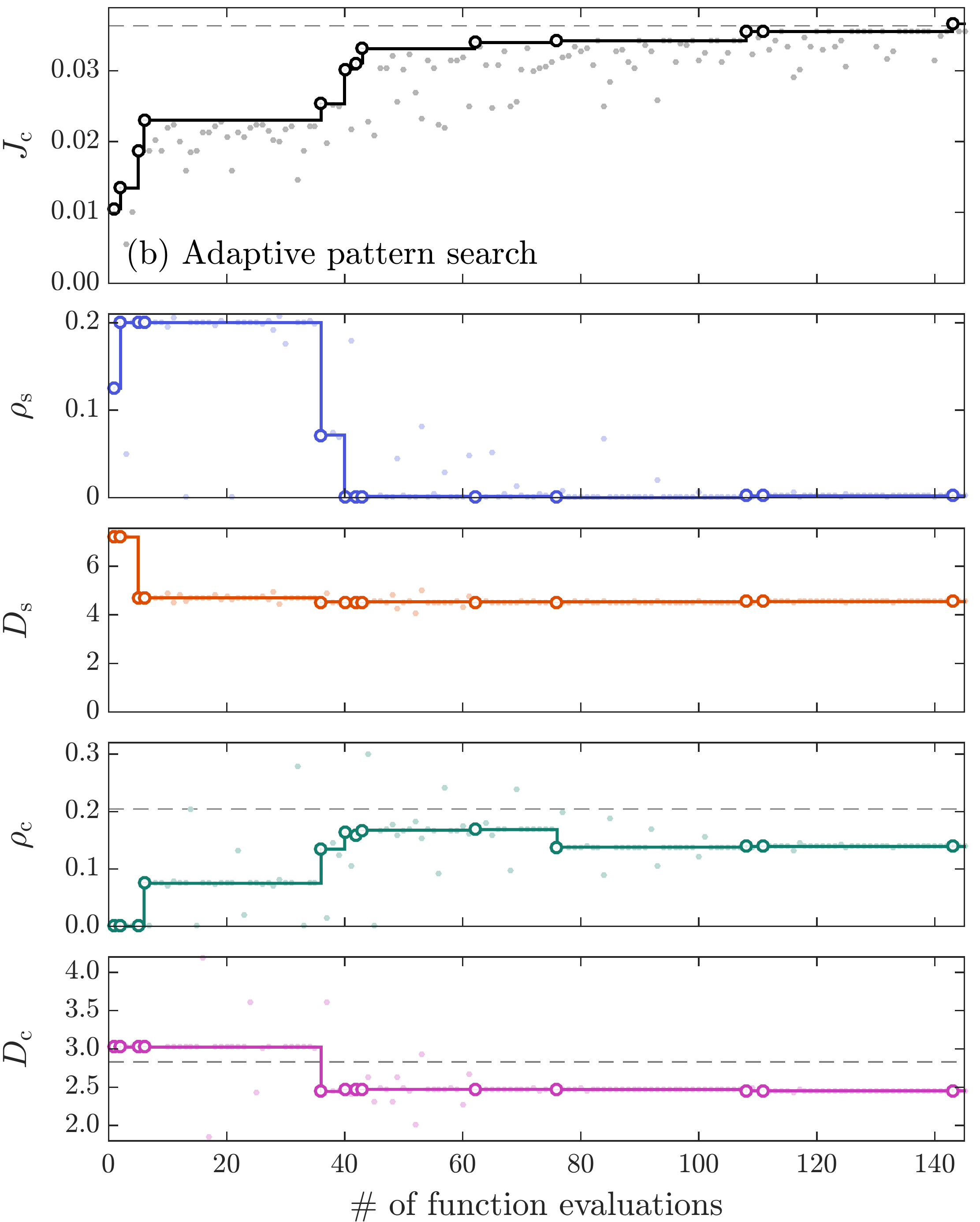} \vspace{0.2cm} \\
	\includegraphics[width=8.8cm]{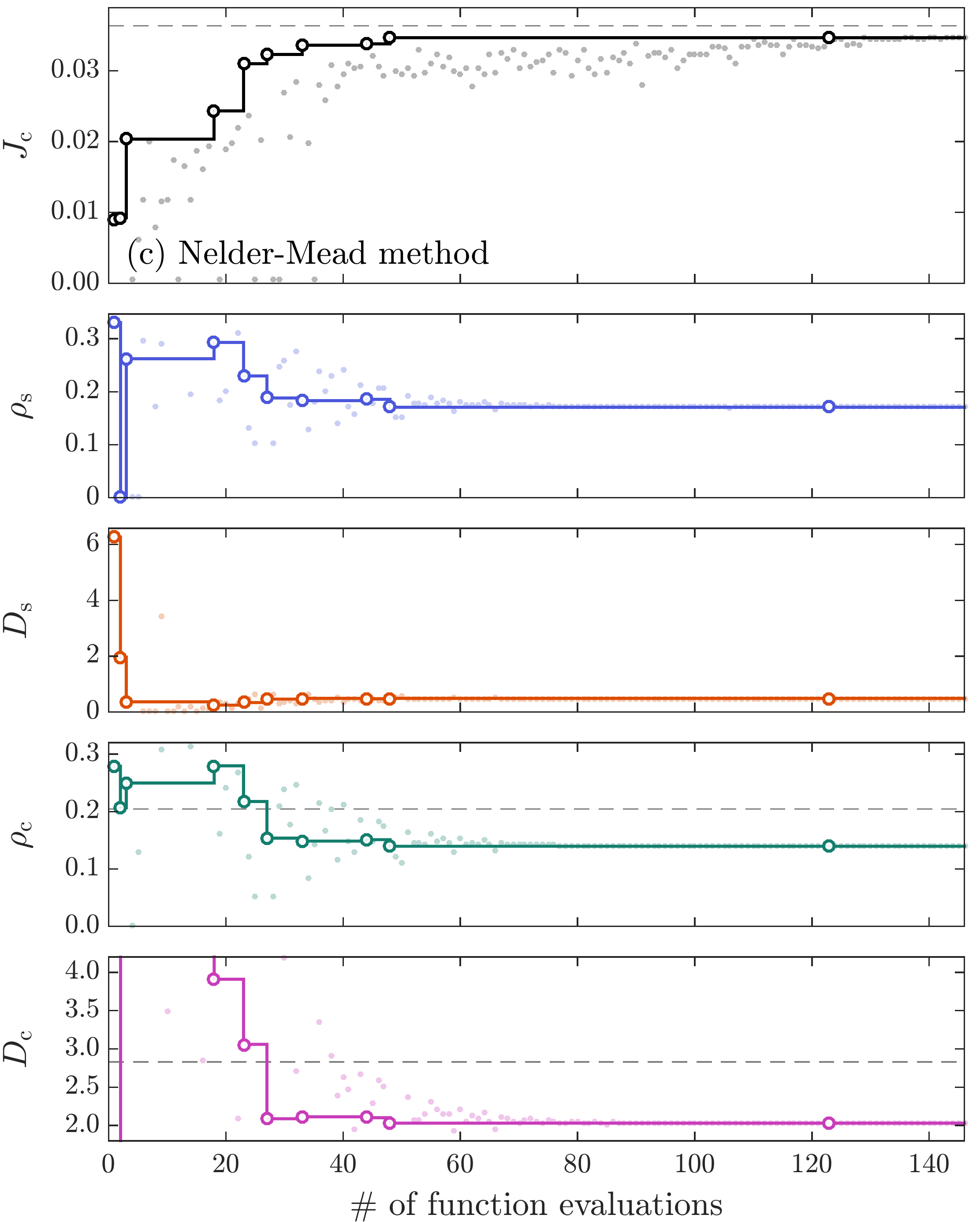} &
	\includegraphics[width=8.8cm]{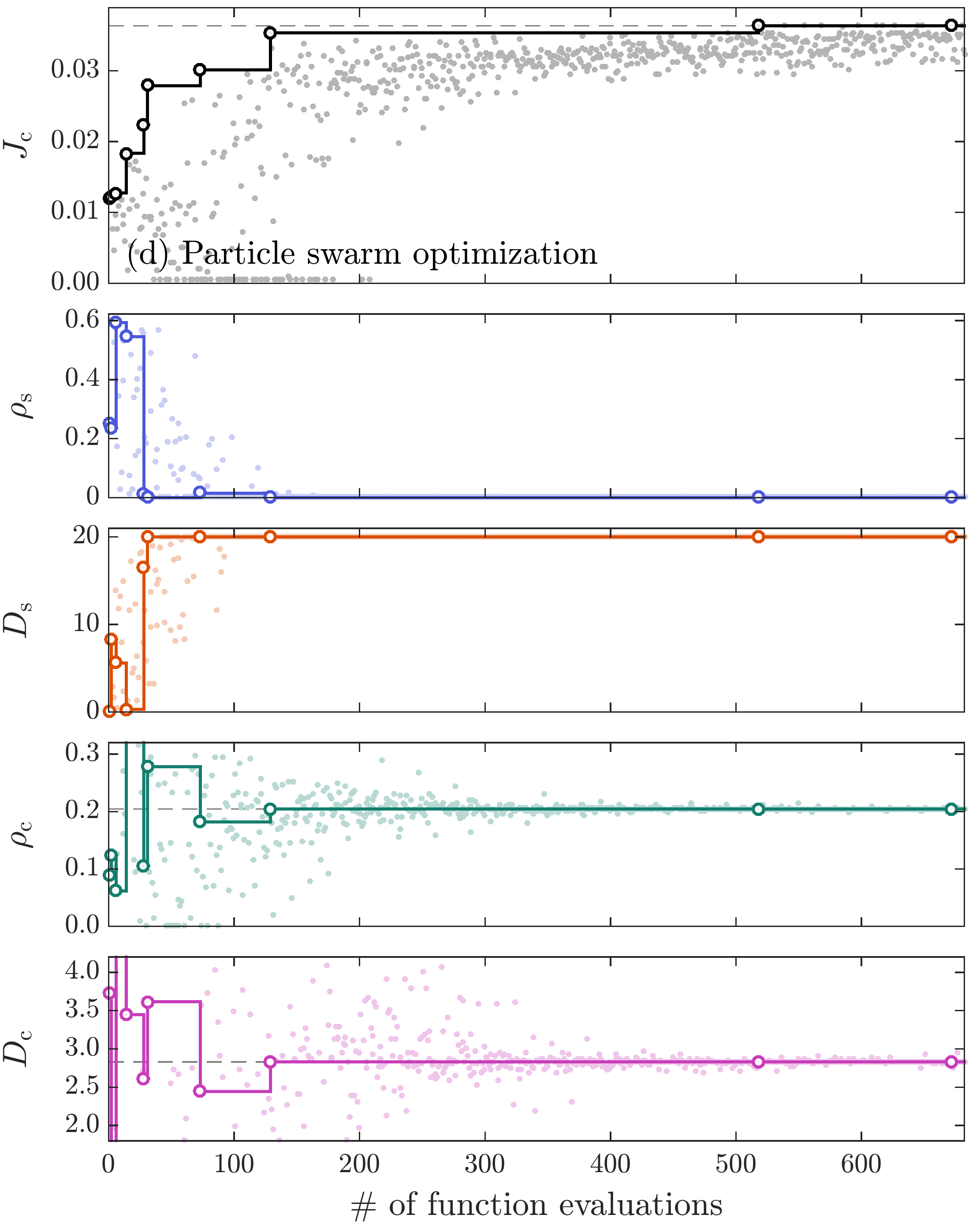}
	\end{tabular}
	\vspace{-0.2cm}
	\caption{
		Optimization procedure for the four-parameter (4D) problem~\eqref{eq:opt4}
		with mixed pinscape containing spherical particles (characterized by volume fraction $\fs$ and diameter in $\Ds$) columnar defects along $z$ axis (volume fraction $\fc$ and diameter in $\Dc$).
		Critical current $\Jc$ and optimization parameters are shown as 
		a function of number of objective function evaluations for
		(a)~pattern search, 
		(b)~adaptive pattern search, 
		(c)~Nelder-Mead method, and 
		(d)~particle swarm optimization.
		PSO and adaptive pattern search converge to the pinscape containing columnar defects only.
	}
\label{fig:d4}
\end{figure*}

To obtain the critical current $\Jc(\x)$, we solve the TDGL equation in the domain of interest with a specified parameter set $\x$. Each evaluation of $\Jc$ is relatively expensive (typically, a few GPU-hours), but can be performed independently for different $\x$. In order to reduce noise in $\Jc$, one can (i)~average it over several realizations of random positions of defects and/or (ii)~increase the system size. Both techniques naturally increase the computation time of $\Jc$. The final value defines the objective function for the optimization problem.

\section{Optimization of the critical current} \label{sec:problem}

The objective function in Eq. \eqref{eq:objfunc} is then used to solve the general optimization problem for the pinscape defined in Eq.~\eqref{eq:opt}. Typical pinscapes with like defects are described by parameter spaces $\Omega$ having $n = 2$ to $8$ dimensions. Here we consider three particular and important cases for $n=2$, $3$, $4$ to analyze the described optimization strategies. For practical applications also the \textit{robustness} of an optimum is important, since e.g. the size and shape of defects cannot be controlled exactly in an experiment. Therefore, the found optimal critical current should be relatively insensitive to small perturbations in the optimal pinscape $\xopt$. In the examples studied here, this condition is fulfilled.

For the simulations needed to obtain the critical current for a given pinscape, we use a three-dimensional superconducting sample with dimensions $64 \times 64 \times 64$ in units of coherence length~$\xi$ with (quasi-)periodic boundary conditions. The external current~$\Jext$ is applied in the $x$-direction, perpendicular to the magnetic field~$B$ in the $z$-direction. All defects are modeled by local variation of the critical temperature, such that defects are regions being in the normal state. To this end the system is discretized by a regular cubic grid with grid resolution of half a coherence length. 
As demonstrated in earlier works,\cite{Sadovskyy:2016a,Sadovskyy:2016b} this resolution is sufficient to capture the involved physical processes correctly. 
A pattern generator then creates the pinscape according to the defining parameter set $\x$, by assigning each grid point either to the superconductor or to the defect, distinguished by a high or low local $\Tc$ value.
Defect regions with low $\Tc$ values typically consist of many connected grid point.

First, we consider randomly placed spherical defects (typically self-assembled inclusions in real sample) in a three-dimensional system. 
All inclusions have the same diameter, such that the parameter space for this problem is defined by the defect density $\fs$ within the sample and their diameter $\Ds$. 
As mentioned before, this set of parameters is not unique, such that the same pinscape could be described by a different $\x$ (e.g., the diameter and total number of defects). 
This two-parameter pinscape problem was considered recently in Ref.~\onlinecite{Koshelev:2016} by sampling of the parameter space, which was possible in that case.

Another example of a two-parameter problem would be a system with two kinds of defects having fixed shape, e.g., a superconductor having columnar defects of fixed diameter imprinted (typically introduced by high-energy heavy-ion irradiation in a real sample), which already had intrinsic (usually chemically grown) nano-rod pinning centers.\cite{Sadovskyy:2016b} 
In this case the parameter space will be defined by the concentration $\fc$ of columnar defects and the concentration $\fn$ of the intrinsic nano-rod defects, i.e., $\x = (\fc, \fn)$. 

For the first problem we consider here, the pinscape is defined by
\begin{subequations}
\begin{equation}
	\x = (\fs, \Ds), \qquad \Ns = \biggl\lfloor \frac{6 L_x L_y L_z \fs}{\pi \Ds^3} \biggr\rfloor,
	\label{eq:opt2}
\end{equation}
where $\Ns$ is the number of identical spherical inclusions to be placed randomly [see Fig.~\subref{fig:pattern_spherical}] and $\lfloor \cdot \rfloor$ denotes the integer rounding function. In the second, three-parameter (3D) optimization problem shown in Fig.~\subref{fig:pattern_spheroidal} we replace the spherical defects by spheroidal defects, which have a different diameter in $z$-direction than in the $xy$-plane, i.e., it is described by two different diameters~--- $\Dxy$ in the $xy$ plane and $\Dz$ in $z$ direction. In this case the pinscape is defined by
\begin{equation}
	\x = (\fe, \Dxy, \Dz), \qquad 
	\Ne = \biggl\lfloor \frac{4 L_x L_y L_z \fe}{\pi \kappa \Dxy^2 \tDz} \biggr\rfloor,
	\label{eq:opt3}
\end{equation} 
where $\tDz = \min\{ \Dz, L_z \}$ and $\kappa = 1 - {\tDz^2}/{3 \Dz^2}$ --- the former definitions take into account  periodic boundary conditions in $z$-direction, when the $z$-diameter grows beyond the simulation cuboid. In the final, four-parameter (4D) problem, we consider two types of defects described by a single parameter and their respective volume fractions, namely spherical and columnar (cylindrical) inclusions, see Fig.~\subref{fig:pattern_spherical_columnar}. The respective pinscape control parameters are then $\fs$, $\Ds$ for spheres and $\fc$, $\Dc$ for columns, i.e.,
\begin{equation} 
	\x = (\fs, \Ds, \fc, \Dc), \qquad 
	\Nc = \biggl\lfloor\frac{4 L_x L_y \fs}{\pi \Dxy^2 } \biggr\rfloor,
	\label{eq:opt4}
\end{equation} 
\end{subequations}
where $\Ns$ was defined in Eq.~\eqref{eq:opt2}. In the following we discuss the results for these three optimization problems as well as the benchmark function optimizations.

\section{Results and discussions} \label{sec:results}

\paragraph*{Discussion of benchmark function optimizations.}
The optimization of parameters for adaptive pattern search and PSO has led to some interesting results. First, the nested PSO algorithm finds many different ``good" choices for the parameters needed. This is to be expected, since each problem is unique, and it would be surprising if multiple sets of parameters did not yield similar results. What was surprising, is the sensitivity of the parameters to small deviations from optima that were obtained on Rastrigin's function. A good example would be in a particular optimization run of adaptive pattern search's parameters for $n = 6$. We found that the set $\{ \mu$, $\sigma$, $k_\mathrm{s}$, $k_\mathrm{u} \} = \{ 6$, $0.3$, $6.59$, $0.29 \}$ was the best for that particular run found with $2 \cdot 10^5$ function evaluations needed. Just changing one of these parameters slightly to $\{ 6$, $0.3$, $6.59$, $0.39\}$, needs $4.25 \cdot 10^5$ function evaluations.

This sensitivity was not observed in the other two benchmark functions, meaning that the sensitivity of these parameters for the optimization routines are highly dependent on the type of function, being particularly sensitive to functions with multiple optima. At the other end of this spectrum, it was observed that many different sets of parameter values led to a minimal amount of evaluations. Essentially, there was not a large difference between the best observed set, which is reported in Tables~\ref{tab:opt_PSO_sphere} and \ref{tab:opt_adaptive_sphere}, and other choices. From this we can conclude that surfaces with uncorrelated dimensional space and with small number of optima are more likely to be insensitive to the choice of parameters and hence do not need to be tuned for the particular problem.

The Rastrigin function is particularly useful in analyzing the effect multiple optima have on local methods. One can find the number of extrema as a function of dimension size $n$. The derivative of the Rastrigin function in each respective coordinate is set to 0 and is given by $x_i + 10 \pi \sin(2 \pi x_i) = 0$. We recall that the bounds for our benchmark functions are $x_i \in [-10,10]$. It is easy to see that due to symmetry, we only need to consider $x_i > 0$. We further note that solutions are not possible for $x_i > 10 \pi$ and for each oscillation period, one obtains two extrema. Since the period is 1, this implies 10 minima (20 extremum which by symmetry has 10 maxima and 10 minima) within the above bounds. Furthermore, symmetry implies 10 more in the negative direction plus the obvious $x_i = 0$ (global) minimum, which results in 21 minima total. It then follows that the number of minima  in our problem for $n$ dimensions is $20^n + 1 \approx 20^n$.

Another aspect of the optimization is how high the probability for a found optimum being the global optimum is. For this we can consider any local method which is run $R$ times, and suppose the method has an equal probability to converge to any of the $m$ minima. We also assume $m \gg 1$, and we want to be $\alpha = 0.99$ sure we have found the global solution. 
This would require $R \approx 4.6 m$ [or $R = O(m)$]. 
Thus the number of local runs should scale approximately linearly with increasing minima. With just $n = 2$ for the Rastrigin function for example would imply that a local method would need to be run $R \approx 2 \cdot 10^3$ times to be 99\% sure we have converged to the global solution. Again, it is important to note that this number assumed each extremum was equally likely, that is, each of their respective basins of attractions were the same size. This assumption in general is not true, but it provides an insight into how large the number of runs of a local method would be needed. For example, suppose a local method converges in 50 evaluations on average to a local extremum. In two dimensions this would require on average close to $4 \cdot 10^5$ total evaluations of the objective function. Compare this with the number needed for PSO. In reality this turns out to be an overestimation, the total number of evaluations for local methods are far lower, but this analysis provides some insight as to why they suffer at this level and how the number of evaluations will scale with increasing number of optimums.

In contrast, PSO and other global methods are more robust to the number of local optima as can be seen by comparing Figs.~\ref{fig:Rosenbrock_function_comparison} and \ref{fig:Rastrigin_function_comparison}, the number of function evaluations is on the same order for both Rosenbrock and Rastrigin functions. A possible reason why PSO behaves similarly for Rosenbrock and Rastrigin is because there is only a "1D path" towards the minimum for Rosenbrock, while the sphere function has a 50\% chance to lower the objective in each direction. Therefore, particles are as likely to become stuck in the Rosenbrock valley as they are in a Rastrigin local minimum.

The PSO optimal parameters also provide some insight into the nature of the function. It helps to first interpret what each parameter does. Increasing $S$ obviously leads to a greater chance of finding the best solution, the cost is an increase in function evaluations. The inertial weight $\omega < 1$ is not a coincidence. As the particles approach the global best, $\mathbf{v}_i \gets \omega \mathbf{v}_i$ and this has a fixed point at $\mathbf{v}_i = 0$. The $\phi_\mathrm{p}$ and $\phi_\mathrm{g}$ are quantities which affect the exploratory nature of the algorithm and the weight associated with the global information that the swarm provides to each individual particle. As $\phi_\mathrm{g} \to 0$ each of the particles behave more independently and their movement becomes more individualized. As $\phi_\mathrm{p} \to 0$, we are essentially telling each particle to only trust the group on the correct direction to go.

Armed with this interpretation, we can understand some of the results from the tables in Sec.~\ref{sec:benchmark_functions}. The sphere function and Rosenbrock both have few minima (sphere has only one, and the Rosenbrock has no more than two for $n \leqslant 7$). Therefore we expect $\phi_\mathrm{g} \geqslant \phi_\mathrm{p}$. However, the results are different in Table~\ref{tab:opt_PSO_Rastrigin}, which shows $\phi_\mathrm{p}$ increasing and $\phi_\mathrm{g}$ decreasing as $n$ (number of optima) increases. This makes sense, with multiple optima, we want the swarm to behave more independently and not get caught among the many local optima. However we don't want $\phi_\mathrm{g} = 0$ or else we would take very long to converge.

\paragraph*{Discussion of critical current optimizations.}
Figures~\ref{fig:d2}, \ref{fig:d3}, and \ref{fig:d4}  show the optimization ``paths'' for all four optimization methods for the 2D [Eq.~\eqref{eq:opt2} with parameter space dimension $n = 2$], 3D  [Eq.~\eqref{eq:opt3} with $n = 3$], and 4D  [Eq.~\eqref{eq:opt4} with $n = 4$] problems, respectively. These plots show the values of the currently best critical current and the associated parameter set $\x$ as function of  number of evaluations of the objective function (solid lines). If a new optimal value is found it is marked by circles, while all other evaluations (with objective values above the current minimum) are marked by gray dots. The optimal values are shown by dashed horizontal lines.
Note, that the pattern search algorithm is not sequential, meaning that a minimum is only calculated after a certain number of evaluations (up to $2n+1$), which defines an iteration of the method. The iterations are marked by vertical dotted lines in the plots. If the number of evaluations per iteration is less that $2n+1$, the remaining objective values were calculated before and are just looked up in a database. Iterations are shaded according to the step sizes in parameter space.
The optimization results are summarized in Table~\ref{tab:opt_Jc}. 
In the 2D case for randomly placed spherical inclusions, the optimal values reproduce the ones of Ref.~\onlinecite{Koshelev:2016} at magnetic field $B = 0.1\Hct$. 
In the 3D case for spheroidal inclusions, the optimal pinscape corresponds to an infinite size in $z$ direction (larger then the system size), i.e., when the spheroids have ``evolved'' to columnar inclusion, and the optimal value of the critical current is increased by $\sim 50\%$. 
In the 4D case, for a combination of the columnar and spherical defects, the 3D result is confirmed, since the volume fraction for the spheres vanish and the best pinscape turns out to be consisting of columnar defects in $z$ direction only. 
The diameter and volume fraction of the best columnar defects in the 3D and 4D cases is accordingly the same.

\begin{table}[tbh]
\centering \normalsize
\begin{tabularx}{0.9\columnwidth}{@{}l *4{R}@{}}
$n$ & $\x = \{ \fs\:\:$ & $\Ds \}$ & $\Jc$\vspace{1.5pt}\\
\hline
2 & 0.22 & $3.5\:\:$ & 0.0235\vspace{1mm}\\
\end{tabularx}
\begin{tabularx}{0.9\columnwidth}{@{}l *5{R}@{}}
$n$ & $\x = \{ \f\:\:\:$ & $\Dxy$ & $\Dz \}$ & $\Jc$\vspace{1.5pt}\\
\hline
3 & 0.20 & 3.0 & $\infty\:\:$ & 0.035\vspace{1mm}\\
\end{tabularx}
\begin{tabularx}{0.9\columnwidth}{@{}l *6{R}@{}}
$n$ & $\x = \{ \fs\:\:$ & $\Ds$ & $\fc$ & $\Dc \}$ & $\Jc$\vspace{1.5pt}\\
\hline
4 & 0.00 & --- & 0.20 & $3.0\:\:$ & 0.035 \\
\end{tabularx}
\caption{Optimal parameters of the pinscape for maximal critical current in the superconductor.}
\label{tab:opt_Jc}
\end{table}

As mentioned before, in order to obtain a good critical current value for randomly placed inclusions, either a large number of random realizations or much larger systems are needed. While the latter is easier to run, we chose the former for performance reasons, such that a single objective function evaluation requires averaging the critical current over many pinscapes, described by the same $\x$. However, even with a large number of realizations, the critical current as a function of $\x$ can still be quite noisy. In the worst case this leads to local methods failing to reach the optimal solution. In contrast to that, global methods like PSO are much more resistant to noise in the objective function.

First, let us analyze the two parameter optimization more closely (see Fig.~\ref{fig:d2}). All optimization methods converge to the same optimal pinscape. However, the optimization ``paths'' are quite different as one could expect: although all methods reach a critical current value close to the optimum after about 10--20 evaluations~--- indicating a robust, relatively flat peak in parameter space, the actual optimal configuration takes quite a while to be reached: It takes about twice as many evaluations for the local methods and an order of magnitude more for the global PSO method. Note, that defining an appropriate exit criterion for the optimization can be challenging, since the improvement in the critical current can be marginal, while the configuration can still change noticeably, as is clearly seen in the Nelder-Mead path.

Increasing the parameter space to three dimensions, (Fig.~\ref{fig:d3}), shows that all methods need more evaluations as expected (scaling roughly linear with the dimension) and again the local methods reach a close to optimal critical current significantly faster than PSO. However, one can see that the local methods already start to have convergence problems, which is most likely a result of the noise in the objective function. The figure shows that the adaptive pattern search is a bit more efficient dealing with this noise than simple coordinate decent.  However, in this case all methods still reach a configuration close to the optimal one.

An obvious observation is that increasing the parameter space, cannot lower the optimal value of the objective function. Indeed, the chosen examples are in a sense supersets of each other with higher dimensionality of $\Omega$, such that the best $\Jc$ value either increases or remains the same. When going over from the 3D to 4D optimization problem, we kept the result for the optimal $z$-diameter of the defects, i.e., that the optimal defects are cylinders, but added back spherical defects to the system. In the 4D case one could then imagine three different scenarios for the global optimal configuration: coexistence of both defect types, only cylindrical defects remain, or only spherical are optimal. Due to the results of the 3D problem, the latter is rather unlikely, since vortices and columnar defects are well aligned and the pinning potential of the defects are therefore ``optimal.''

As seen from Fig.~\ref{fig:d4}, the local methods converge to different optimal solutions. The coexistence scenario appears to be a good pinscape. However, the largest $\Jc$ value is still below the cylinder-only solution. One can expect that the optimal configuration is dependent on the angle between applied magnetic field and the main axis of the cylindrical defects. If they deviate from the parallel setup we studied here, the coexistence of spherical and columnar defects might be optimal, but it is clear that at a 90$^\circ$ angle between field and columns only spherical defects would ``survive'' in an optimization problem (for isotropic superconductors).

When we compare our results for the $\Jc$ optimization to the studied test functions before, we can conclude, that the $\Jc$ surface for the chosen parameters is roughly uncorrelated, since the number of evaluations required for adaptive pattern search and pattern search to converge are nearly the same.

Overall, we see that the PSO algorithm works reliable in all cases, at the cost of about 10 times more evaluations needed to reach the optimum, where often many evaluations, on the order $10^2$--$10^3$, are needed for a marginal gain in the objective function ($< 1\%$). The local methods start to fail in higher dimensional parameter spaces, most likely due to noise in the objective function. As mentioned before, combining the local searches with a global method can potentially reduce the overall number of evaluations. Another way to mitigate this problem is to use more realizations or larger system sizes, which should smooth out the error from the random placement of defects. 

Another benefit of using a global method such as PSO, is that it consistently finds better solution than the local methods. In higher dimensional spaces $n \gtrsim 4$, taking a larger swarm size $S$ becomes necessary to ensure convergence. The local methods, however, converge faster. But in any case there is no guarantee that the found configuration is the optimum. One could only estimate a probability that the found result is close to the optimal one.

Finally, we remark that larger oscillations in the optimization paths for the local methods in higher dimensions suggest a flatter global maximum of $\Jc$ or a more robust critical current when one deviates from the optimal configuring. This is particularly important for practical applications.

\section{Conclusions} \label{sec:conclusions}

We have tested various methods for optimization problems and determined that particle swarm optimization and adaptive pattern search performed the best on the benchmark functions as well as in the physical optimization problem presented. Particle swarm optimization in particular does a relatively good job at handling the noisy surface of the objective function, but takes a long time to converge, where a significant amount of time is used for marginal improvement. This is a result of the difficulty to define an exit criterion. On the other side, adaptive pattern search can get caught in local extrema, which are either physical or a result of the noise. However, the number of evaluations are often much less (on the order of 7--10 times). As mentioned before, a multi-level local method could be used to mitigate this, which should make adaptive pattern search competitive with particle swarm optimization. Although we presented only some results for the pinscape optimization in superconductors, our studies of different optimization strategies can be important for a variety of different physical systems where the evaluation of the objective function is expensive, e.g., requiring first principle calculations or molecular dynamics simulations.

\acknowledgements

This work was supported by the Scientific Discovery through Advanced Computing (SciDAC) program funded by U.S. Department of Energy, Office of Science, Advanced Scientific Computing Research and Basic Energy Science, Division of Materials Science and Engineering. For the computational work we used resources of the Argonne Leadership Computing Facility (DOE Office of Science User Facility supported under Contract DE-AC02-06CH11357) and the GPU cluster GAEA at Northern Illinois University.

\bibliographystyle{apsrev4-1-titles}

\begin{thebibliography}{34}%
\makeatletter
\providecommand \@ifxundefined [1]{%
 \@ifx{#1\undefined}
}%
\providecommand \@ifnum [1]{%
 \ifnum #1\expandafter \@firstoftwo
 \else \expandafter \@secondoftwo
 \fi
}%
\providecommand \@ifx [1]{%
 \ifx #1\expandafter \@firstoftwo
 \else \expandafter \@secondoftwo
 \fi
}%
\providecommand \natexlab [1]{#1}%
\providecommand \enquote  [1]{``#1''}%
\providecommand \bibnamefont  [1]{#1}%
\providecommand \bibfnamefont [1]{#1}%
\providecommand \citenamefont [1]{#1}%
\providecommand \href@noop [0]{\@secondoftwo}%
\providecommand \href [0]{\begingroup \@sanitize@url \@href}%
\providecommand \@href[1]{\@@startlink{#1}\@@href}%
\providecommand \@@href[1]{\endgroup#1\@@endlink}%
\providecommand \@sanitize@url [0]{\catcode `\\12\catcode `\$12\catcode
  `\&12\catcode `\#12\catcode `\^12\catcode `\_12\catcode `\%12\relax}%
\providecommand \@@startlink[1]{}%
\providecommand \@@endlink[0]{}%
\providecommand \url  [0]{\begingroup\@sanitize@url \@url }%
\providecommand \@url [1]{\endgroup\@href {#1}{\urlprefix }}%
\providecommand \urlprefix  [0]{URL }%
\providecommand \Eprint [0]{\href }%
\providecommand \doibase [0]{http://dx.doi.org/}%
\providecommand \selectlanguage [0]{\@gobble}%
\providecommand \bibinfo  [0]{\@secondoftwo}%
\providecommand \bibfield  [0]{\@secondoftwo}%
\providecommand \translation [1]{[#1]}%
\providecommand \BibitemOpen [0]{}%
\providecommand \bibitemStop [0]{}%
\providecommand \bibitemNoStop [0]{.\EOS\space}%
\providecommand \EOS [0]{\spacefactor3000\relax}%
\providecommand \BibitemShut  [1]{\csname bibitem#1\endcsname}%
\let\auto@bib@innerbib\@empty
\bibitem [{\citenamefont {Kwok}\ \emph {et~al.}(2016)\citenamefont {Kwok},
  \citenamefont {Welp}, \citenamefont {Glatz}, \citenamefont {Koshelev},
  \citenamefont {Kihlstrom},\ and\ \citenamefont {Crabtree}}]{Kwok:2016}%
  \BibitemOpen
  \bibfield  {author} {\bibinfo {author} {\bibfnamefont {W.-K.}\ \bibnamefont
  {Kwok}}, \bibinfo {author} {\bibfnamefont {U.}~\bibnamefont {Welp}}, \bibinfo
  {author} {\bibfnamefont {A.}~\bibnamefont {Glatz}}, \bibinfo {author}
  {\bibfnamefont {A.~E.}\ \bibnamefont {Koshelev}}, \bibinfo {author}
  {\bibfnamefont {K.~J.}\ \bibnamefont {Kihlstrom}}, \ and\ \bibinfo {author}
  {\bibfnamefont {G.~W.}\ \bibnamefont {Crabtree}},\ }\bibfield  {title} {\emph
  {\bibinfo {title} {Vortices in high-performance high-temperature
  superconductors}},\ }\href {\doibase 10.1088/0034-4885/79/11/116501}
  {\bibfield  {journal} {\bibinfo  {journal} {Rep. Prog. Phys.}\ }\textbf
  {\bibinfo {volume} {79}},\ \bibinfo {pages} {116501} (\bibinfo {year}
  {2016})}\BibitemShut {NoStop}%
\bibitem [{\citenamefont {Holesinger}\ \emph {et~al.}(2008)\citenamefont
  {Holesinger}, \citenamefont {Civale}, \citenamefont {Maiorov}, \citenamefont
  {Feldmann}, \citenamefont {Coulter}, \citenamefont {Miller}, \citenamefont
  {Maroni}, \citenamefont {Chen}, \citenamefont {Larbalestier}, \citenamefont
  {Feenstra} \emph {et~al.}}]{Holesinger:2008}%
  \BibitemOpen
  \bibfield  {author} {\bibinfo {author} {\bibfnamefont {T.~G.}\ \bibnamefont
  {Holesinger}}, \bibinfo {author} {\bibfnamefont {L.}~\bibnamefont {Civale}},
  \bibinfo {author} {\bibfnamefont {B.}~\bibnamefont {Maiorov}}, \bibinfo
  {author} {\bibfnamefont {D.~M.}\ \bibnamefont {Feldmann}}, \bibinfo {author}
  {\bibfnamefont {J.~Y.}\ \bibnamefont {Coulter}}, \bibinfo {author}
  {\bibfnamefont {D.~J.}\ \bibnamefont {Miller}}, \bibinfo {author}
  {\bibfnamefont {V.~A.}\ \bibnamefont {Maroni}}, \bibinfo {author}
  {\bibfnamefont {Z.}~\bibnamefont {Chen}}, \bibinfo {author} {\bibfnamefont
  {D.~C.}\ \bibnamefont {Larbalestier}}, \bibinfo {author} {\bibfnamefont
  {R.}~\bibnamefont {Feenstra}},  \emph {et~al.},\ }\bibfield  {title} {\emph
  {\bibinfo {title} {Progress in nanoengineered microstructures for tunable
  high-current, high-temperature superconducting wires}},\ }\href {\doibase
  10.1002/adma.200700919} {\bibfield  {journal} {\bibinfo  {journal} {Adv.
  Mat.}\ }\textbf {\bibinfo {volume} {20}},\ \bibinfo {pages} {391} (\bibinfo
  {year} {2008})}\BibitemShut {NoStop}%
\bibitem [{\citenamefont {Malozemoff}(2012)}]{Malozemoff:2012}%
  \BibitemOpen
  \bibfield  {author} {\bibinfo {author} {\bibfnamefont {A.}~\bibnamefont
  {Malozemoff}},\ }\bibfield  {title} {\emph {\bibinfo {title}
  {Second-generation high-temperature superconductor wires for the electric
  power grid}},\ }\href {\doibase 10.1146/annurev-matsci-100511-100240}
  {\bibfield  {journal} {\bibinfo  {journal} {Annu. Rev. Mater. Res.}\ }\textbf
  {\bibinfo {volume} {42}},\ \bibinfo {pages} {373} (\bibinfo {year}
  {2012})}\BibitemShut {NoStop}%
\bibitem [{\citenamefont {Abrikosov}(1957)}]{Abrikosov:1957}%
  \BibitemOpen
  \bibfield  {author} {\bibinfo {author} {\bibfnamefont {A.~A.}\ \bibnamefont
  {Abrikosov}},\ }\bibfield  {title} {\emph {\bibinfo {title} {The magnetic
  properties of superconducting alloys}},\ }\href {\doibase
  10.1016/0022-3697(57)90083-5} {\bibfield  {journal} {\bibinfo  {journal} {J.
  Phys. Chem. Solids}\ }\textbf {\bibinfo {volume} {2}},\ \bibinfo {pages}
  {199} (\bibinfo {year} {1957})}\BibitemShut {NoStop}%
\bibitem [{\citenamefont {Blatter}\ \emph {et~al.}(1994)\citenamefont
  {Blatter}, \citenamefont {Feigel'man}, \citenamefont {Geshkenbein},
  \citenamefont {Larkin},\ and\ \citenamefont {Vinokur}}]{Blatter:1994}%
  \BibitemOpen
  \bibfield  {author} {\bibinfo {author} {\bibfnamefont {G.}~\bibnamefont
  {Blatter}}, \bibinfo {author} {\bibfnamefont {M.~V.}\ \bibnamefont
  {Feigel'man}}, \bibinfo {author} {\bibfnamefont {V.~B.}\ \bibnamefont
  {Geshkenbein}}, \bibinfo {author} {\bibfnamefont {A.~I.}\ \bibnamefont
  {Larkin}}, \ and\ \bibinfo {author} {\bibfnamefont {V.~M.}\ \bibnamefont
  {Vinokur}},\ }\bibfield  {title} {\emph {\bibinfo {title} {Vortices in
  high-temperature superconductors}},\ }\href {\doibase
  10.1103/RevModPhys.66.1125} {\bibfield  {journal} {\bibinfo  {journal} {Rev.
  Mod. Phys.}\ }\textbf {\bibinfo {volume} {66}},\ \bibinfo {pages} {1125}
  (\bibinfo {year} {1994})}\BibitemShut {NoStop}%
\bibitem [{\citenamefont {Blatter}\ and\ \citenamefont
  {Geshkenbein}(2003)}]{Blatter:2003}%
  \BibitemOpen
  \bibfield  {author} {\bibinfo {author} {\bibfnamefont {G.}~\bibnamefont
  {Blatter}}\ and\ \bibinfo {author} {\bibfnamefont {V.}~\bibnamefont
  {Geshkenbein}},\ }\bibfield  {title} {\emph {\bibinfo {title} {Vortex
  matter}},\ }in\ \href {\doibase 10.1007/978-3-642-55675-3_10} {\emph
  {\bibinfo {booktitle} {The physics of superconductors}}},\ \bibinfo {editor}
  {edited by\ \bibinfo {editor} {\bibfnamefont {K.}~\bibnamefont {Bennemann}}\
  and\ \bibinfo {editor} {\bibfnamefont {J.}~\bibnamefont {Ketterson}}}\
  (\bibinfo  {publisher} {Springer Berlin Heidelberg},\ \bibinfo {year}
  {2003})\ pp.\ \bibinfo {pages} {725--936}\BibitemShut {NoStop}%
\bibitem [{\citenamefont {Ortalan}\ \emph {et~al.}(2009)\citenamefont
  {Ortalan}, \citenamefont {Herrera}, \citenamefont {Rupich},\ and\
  \citenamefont {Browning}}]{Ortalan:2009}%
  \BibitemOpen
  \bibfield  {author} {\bibinfo {author} {\bibfnamefont {V.}~\bibnamefont
  {Ortalan}}, \bibinfo {author} {\bibfnamefont {M.}~\bibnamefont {Herrera}},
  \bibinfo {author} {\bibfnamefont {M.~W.}\ \bibnamefont {Rupich}}, \ and\
  \bibinfo {author} {\bibfnamefont {N.~D.}\ \bibnamefont {Browning}},\
  }\bibfield  {title} {\emph {\bibinfo {title} {Three dimensional analyses of
  flux pinning centers in {Dy}-doped {YBa$_2$Cu$_3$O$_{7-x}$} coated
  superconductors by {STEM} tomography}},\ }\href {\doibase
  10.1016/j.physc.2009.08.012} {\bibfield  {journal} {\bibinfo  {journal}
  {Phys. C}\ }\textbf {\bibinfo {volume} {469}},\ \bibinfo {pages} {2052}
  (\bibinfo {year} {2009})}\BibitemShut {NoStop}%
\bibitem [{\citenamefont {Sadovskyy}\ \emph
  {et~al.}(2016{\natexlab{a}})\citenamefont {Sadovskyy}, \citenamefont
  {Koshelev}, \citenamefont {Glatz}, \citenamefont {Ortalan}, \citenamefont
  {Rupich},\ and\ \citenamefont {Leroux}}]{Sadovskyy:2016a}%
  \BibitemOpen
  \bibfield  {author} {\bibinfo {author} {\bibfnamefont {I.~A.}\ \bibnamefont
  {Sadovskyy}}, \bibinfo {author} {\bibfnamefont {A.~E.}\ \bibnamefont
  {Koshelev}}, \bibinfo {author} {\bibfnamefont {A.}~\bibnamefont {Glatz}},
  \bibinfo {author} {\bibfnamefont {V.}~\bibnamefont {Ortalan}}, \bibinfo
  {author} {\bibfnamefont {M.~W.}\ \bibnamefont {Rupich}}, \ and\ \bibinfo
  {author} {\bibfnamefont {M.}~\bibnamefont {Leroux}},\ }\bibfield  {title}
  {\emph {\bibinfo {title} {Simulation of the vortex dynamics in a real pinning
  landscape of {YBa$_2$Cu$_3$O$_{7-\delta}$} coated conductors}},\ }\href
  {\doibase 10.1103/PhysRevApplied.5.014011} {\bibfield  {journal} {\bibinfo
  {journal} {Phys. Rev. Applied}\ }\textbf {\bibinfo {volume} {5}},\ \bibinfo
  {pages} {014011} (\bibinfo {year} {2016}{\natexlab{a}})}\BibitemShut
  {NoStop}%
\bibitem [{\citenamefont {Sadovskyy}\ \emph
  {et~al.}(2016{\natexlab{b}})\citenamefont {Sadovskyy}, \citenamefont {Jia},
  \citenamefont {Leroux}, \citenamefont {Kwon}, \citenamefont {Hu},
  \citenamefont {Fang}, \citenamefont {Chaparro}, \citenamefont {Zhu},
  \citenamefont {Welp}, \citenamefont {Zuo}, \citenamefont {Zhang},
  \citenamefont {Nakasaki}, \citenamefont {Selvamanickam}, \citenamefont
  {Crabtree}, \citenamefont {Koshelev}, \citenamefont {Glatz},\ and\
  \citenamefont {Kwok}}]{Sadovskyy:2016b}%
  \BibitemOpen
  \bibfield  {author} {\bibinfo {author} {\bibfnamefont {I.~A.}\ \bibnamefont
  {Sadovskyy}}, \bibinfo {author} {\bibfnamefont {Y.}~\bibnamefont {Jia}},
  \bibinfo {author} {\bibfnamefont {M.}~\bibnamefont {Leroux}}, \bibinfo
  {author} {\bibfnamefont {J.}~\bibnamefont {Kwon}}, \bibinfo {author}
  {\bibfnamefont {H.}~\bibnamefont {Hu}}, \bibinfo {author} {\bibfnamefont
  {L.}~\bibnamefont {Fang}}, \bibinfo {author} {\bibfnamefont {C.}~\bibnamefont
  {Chaparro}}, \bibinfo {author} {\bibfnamefont {S.}~\bibnamefont {Zhu}},
  \bibinfo {author} {\bibfnamefont {U.}~\bibnamefont {Welp}}, \bibinfo {author}
  {\bibfnamefont {J.-M.}\ \bibnamefont {Zuo}}, \bibinfo {author} {\bibfnamefont
  {Y.}~\bibnamefont {Zhang}}, \bibinfo {author} {\bibfnamefont
  {R.}~\bibnamefont {Nakasaki}}, \bibinfo {author} {\bibfnamefont
  {V.}~\bibnamefont {Selvamanickam}}, \bibinfo {author} {\bibfnamefont {G.~W.}\
  \bibnamefont {Crabtree}}, \bibinfo {author} {\bibfnamefont {A.~E.}\
  \bibnamefont {Koshelev}}, \bibinfo {author} {\bibfnamefont {A.}~\bibnamefont
  {Glatz}}, \ and\ \bibinfo {author} {\bibfnamefont {W.-K.}\ \bibnamefont
  {Kwok}},\ }\bibfield  {title} {\emph {\bibinfo {title} {Toward
  superconducting critical current by design}},\ }\href {\doibase
  10.1002/adma.201600602} {\bibfield  {journal} {\bibinfo  {journal} {Adv.
  Mat.}\ }\textbf {\bibinfo {volume} {28}},\ \bibinfo {pages} {4593} (\bibinfo
  {year} {2016}{\natexlab{b}})}\BibitemShut {NoStop}%
\bibitem [{\citenamefont {Leroux}\ \emph {et~al.}(2015)\citenamefont {Leroux},
  \citenamefont {Kihlstrom}, \citenamefont {Holleis}, \citenamefont {Rupich},
  \citenamefont {Sathyamurthy}, \citenamefont {Fleshler}, \citenamefont
  {Sheng}, \citenamefont {Miller}, \citenamefont {Eley}, \citenamefont
  {Civale}, \citenamefont {Kayani}, \citenamefont {Niraula}, \citenamefont
  {Welp},\ and\ \citenamefont {Kwok}}]{Leroux:2015}%
  \BibitemOpen
  \bibfield  {author} {\bibinfo {author} {\bibfnamefont {M.}~\bibnamefont
  {Leroux}}, \bibinfo {author} {\bibfnamefont {K.~J.}\ \bibnamefont
  {Kihlstrom}}, \bibinfo {author} {\bibfnamefont {S.}~\bibnamefont {Holleis}},
  \bibinfo {author} {\bibfnamefont {M.~W.}\ \bibnamefont {Rupich}}, \bibinfo
  {author} {\bibfnamefont {S.}~\bibnamefont {Sathyamurthy}}, \bibinfo {author}
  {\bibfnamefont {S.}~\bibnamefont {Fleshler}}, \bibinfo {author}
  {\bibfnamefont {H.}~\bibnamefont {Sheng}}, \bibinfo {author} {\bibfnamefont
  {D.~J.}\ \bibnamefont {Miller}}, \bibinfo {author} {\bibfnamefont
  {S.}~\bibnamefont {Eley}}, \bibinfo {author} {\bibfnamefont {L.}~\bibnamefont
  {Civale}}, \bibinfo {author} {\bibfnamefont {A.}~\bibnamefont {Kayani}},
  \bibinfo {author} {\bibfnamefont {P.~M.}\ \bibnamefont {Niraula}}, \bibinfo
  {author} {\bibfnamefont {U.}~\bibnamefont {Welp}}, \ and\ \bibinfo {author}
  {\bibfnamefont {W.-K.}\ \bibnamefont {Kwok}},\ }\bibfield  {title} {\emph
  {\bibinfo {title} {Rapid doubling of the critical current of
  {YBa$_2$Cu$_3$O$_{7-\delta}$} coated conductors for viable high-speed
  industrial processing}},\ }\href {\doibase 10.1063/1.4935335} {\bibfield
  {journal} {\bibinfo  {journal} {Appl. Phys. Lett.}\ }\textbf {\bibinfo
  {volume} {107}},\ \bibinfo {pages} {192601} (\bibinfo {year}
  {2015})}\BibitemShut {NoStop}%
\bibitem [{\citenamefont {Eley}\ \emph {et~al.}(2017)\citenamefont {Eley},
  \citenamefont {Leroux}, \citenamefont {Rupich}, \citenamefont {Miller},
  \citenamefont {Sheng}, \citenamefont {Niraula}, \citenamefont {Kayani},
  \citenamefont {Welp}, \citenamefont {Kwok},\ and\ \citenamefont
  {Civale}}]{Eley:2017}%
  \BibitemOpen
  \bibfield  {author} {\bibinfo {author} {\bibfnamefont {S.}~\bibnamefont
  {Eley}}, \bibinfo {author} {\bibfnamefont {M.}~\bibnamefont {Leroux}},
  \bibinfo {author} {\bibfnamefont {M.~W.}\ \bibnamefont {Rupich}}, \bibinfo
  {author} {\bibfnamefont {D.~J.}\ \bibnamefont {Miller}}, \bibinfo {author}
  {\bibfnamefont {H.}~\bibnamefont {Sheng}}, \bibinfo {author} {\bibfnamefont
  {P.~M.}\ \bibnamefont {Niraula}}, \bibinfo {author} {\bibfnamefont
  {A.}~\bibnamefont {Kayani}}, \bibinfo {author} {\bibfnamefont
  {U.}~\bibnamefont {Welp}}, \bibinfo {author} {\bibfnamefont {W.-K.}\
  \bibnamefont {Kwok}}, \ and\ \bibinfo {author} {\bibfnamefont
  {L.}~\bibnamefont {Civale}},\ }\bibfield  {title} {\emph {\bibinfo {title}
  {Decoupling and tuning competing effects of different types of defects on
  flux creep in irradiated {YBa$_2$Cu$_3$O$_{7-\delta}$} coated conductors}},\
  }\href {\doibase 10.1088/0953-2048/30/1/015010} {\bibfield  {journal}
  {\bibinfo  {journal} {Supercond. Sci. Tech.}\ }\textbf {\bibinfo {volume}
  {30}},\ \bibinfo {pages} {015010} (\bibinfo {year} {2017})}\BibitemShut
  {NoStop}%
\bibitem [{\citenamefont {Sadovskyy}\ \emph {et~al.}(2015)\citenamefont
  {Sadovskyy}, \citenamefont {Koshelev}, \citenamefont {Phillips},
  \citenamefont {Karpeyev},\ and\ \citenamefont {Glatz}}]{Sadovskyy:2015a}%
  \BibitemOpen
  \bibfield  {author} {\bibinfo {author} {\bibfnamefont {I.~A.}\ \bibnamefont
  {Sadovskyy}}, \bibinfo {author} {\bibfnamefont {A.~E.}\ \bibnamefont
  {Koshelev}}, \bibinfo {author} {\bibfnamefont {C.~L.}\ \bibnamefont
  {Phillips}}, \bibinfo {author} {\bibfnamefont {D.~A.}\ \bibnamefont
  {Karpeyev}}, \ and\ \bibinfo {author} {\bibfnamefont {A.}~\bibnamefont
  {Glatz}},\ }\bibfield  {title} {\emph {\bibinfo {title} {Stable large-scale
  solver for {Ginzburg-Landau} equations for superconductors}},\ }\href
  {\doibase 10.1016/j.jcp.2015.04.002} {\bibfield  {journal} {\bibinfo
  {journal} {J. Comp. Phys.}\ }\textbf {\bibinfo {volume} {294}},\ \bibinfo
  {pages} {639} (\bibinfo {year} {2015})}\BibitemShut {NoStop}%
\bibitem [{\citenamefont {Vodolazov}(2013)}]{Vodolazov:2013}%
  \BibitemOpen
  \bibfield  {author} {\bibinfo {author} {\bibfnamefont {D.~Y.}\ \bibnamefont
  {Vodolazov}},\ }\bibfield  {title} {\emph {\bibinfo {title} {Vortex-induced
  negative magnetoresistance and peak effect in narrow superconducting
  films}},\ }\href {\doibase 10.1103/PhysRevB.88.014525} {\bibfield  {journal}
  {\bibinfo  {journal} {Phys. Rev. B}\ }\textbf {\bibinfo {volume} {88}},\
  \bibinfo {pages} {014525} (\bibinfo {year} {2013})}\BibitemShut {NoStop}%
\bibitem [{\citenamefont {Berdiyorov}\ \emph
  {et~al.}(2006{\natexlab{a}})\citenamefont {Berdiyorov}, \citenamefont
  {Milo\v{s}evi\'{c}},\ and\ \citenamefont {Peeters}}]{Berdiyorov:2006}%
  \BibitemOpen
  \bibfield  {author} {\bibinfo {author} {\bibfnamefont {G.~R.}\ \bibnamefont
  {Berdiyorov}}, \bibinfo {author} {\bibfnamefont {M.~V.}\ \bibnamefont
  {Milo\v{s}evi\'{c}}}, \ and\ \bibinfo {author} {\bibfnamefont {F.~M.}\
  \bibnamefont {Peeters}},\ }\bibfield  {title} {\emph {\bibinfo {title} {Novel
  commensurability effects in superconducting films with antidot arrays}},\
  }\href {\doibase 10.1103/PhysRevLett.96.207001} {\bibfield  {journal}
  {\bibinfo  {journal} {Phys. Rev. Lett.}\ }\textbf {\bibinfo {volume} {96}},\
  \bibinfo {pages} {207001} (\bibinfo {year} {2006}{\natexlab{a}})}\BibitemShut
  {NoStop}%
\bibitem [{\citenamefont {Berdiyorov}\ \emph
  {et~al.}(2006{\natexlab{b}})\citenamefont {Berdiyorov}, \citenamefont
  {Milo\v{s}evi\'{c}},\ and\ \citenamefont {Peeters}}]{Berdiyorov:2006b}%
  \BibitemOpen
  \bibfield  {author} {\bibinfo {author} {\bibfnamefont {G.~R.}\ \bibnamefont
  {Berdiyorov}}, \bibinfo {author} {\bibfnamefont {M.~V.}\ \bibnamefont
  {Milo\v{s}evi\'{c}}}, \ and\ \bibinfo {author} {\bibfnamefont {F.~M.}\
  \bibnamefont {Peeters}},\ }\bibfield  {title} {\emph {\bibinfo {title}
  {Superconducting films with antidot arrays~--- novel behavior of the critical
  current}},\ }\href {\doibase 10.1209/epl/i2006-10013-1} {\bibfield  {journal}
  {\bibinfo  {journal} {Europhys. Lett.}\ }\textbf {\bibinfo {volume} {74}},\
  \bibinfo {pages} {493} (\bibinfo {year} {2006}{\natexlab{b}})}\BibitemShut
  {NoStop}%
\bibitem [{\citenamefont {Sadovskyy}\ \emph {et~al.}(2017)\citenamefont
  {Sadovskyy}, \citenamefont {Wang}, \citenamefont {Xiao}, \citenamefont
  {Kwok},\ and\ \citenamefont {Glatz}}]{Sadovskyy:2017}%
  \BibitemOpen
  \bibfield  {author} {\bibinfo {author} {\bibfnamefont {I.~A.}\ \bibnamefont
  {Sadovskyy}}, \bibinfo {author} {\bibfnamefont {Y.~L.}\ \bibnamefont {Wang}},
  \bibinfo {author} {\bibfnamefont {Z.-L.}\ \bibnamefont {Xiao}}, \bibinfo
  {author} {\bibfnamefont {W.-K.}\ \bibnamefont {Kwok}}, \ and\ \bibinfo
  {author} {\bibfnamefont {A.}~\bibnamefont {Glatz}},\ }\bibfield  {title}
  {\emph {\bibinfo {title} {Effect of hexagonal patterned arrays and defect
  geometry on the critical current of superconducting films}},\ }\href
  {\doibase 10.1103/PhysRevB.95.075303} {\bibfield  {journal} {\bibinfo
  {journal} {Phys. Rev. B}\ }\textbf {\bibinfo {volume} {95}},\ \bibinfo
  {pages} {075303} (\bibinfo {year} {2017})}\BibitemShut {NoStop}%
\bibitem [{\citenamefont {Hooke}\ and\ \citenamefont
  {Jeeves}(1961)}]{Hooke:1961}%
  \BibitemOpen
  \bibfield  {author} {\bibinfo {author} {\bibfnamefont {R.}~\bibnamefont
  {Hooke}}\ and\ \bibinfo {author} {\bibfnamefont {T.~A.}\ \bibnamefont
  {Jeeves}},\ }\bibfield  {title} {\emph {\bibinfo {title} {``{Direct} search''
  solution of numerical and statistical problems}},\ }\href {\doibase
  10.1145/321062.321069} {\bibfield  {journal} {\bibinfo  {journal} {J. ACM}\
  }\textbf {\bibinfo {volume} {8}},\ \bibinfo {pages} {212} (\bibinfo {year}
  {1961})}\BibitemShut {NoStop}%
\bibitem [{\citenamefont {Torczon}(1997)}]{Torczon:1997}%
  \BibitemOpen
  \bibfield  {author} {\bibinfo {author} {\bibfnamefont {V.}~\bibnamefont
  {Torczon}},\ }\bibfield  {title} {\emph {\bibinfo {title} {On the convergence
  of pattern search algorithms}},\ }\href {\doibase 10.1137/S1052623493250780}
  {\bibfield  {journal} {\bibinfo  {journal} {SIAM J. Optim.}\ }\textbf
  {\bibinfo {volume} {7}},\ \bibinfo {pages} {1} (\bibinfo {year}
  {1997})}\BibitemShut {NoStop}%
\bibitem [{\citenamefont {Audet}\ and\ \citenamefont
  {Dennis~Jr}(2002)}]{Audet:2002}%
  \BibitemOpen
  \bibfield  {author} {\bibinfo {author} {\bibfnamefont {C.}~\bibnamefont
  {Audet}}\ and\ \bibinfo {author} {\bibfnamefont {J.~E.}\ \bibnamefont
  {Dennis~Jr}},\ }\bibfield  {title} {\emph {\bibinfo {title} {Analysis of
  generalized pattern searches}},\ }\href {\doibase 10.1137/S1052623400378742}
  {\bibfield  {journal} {\bibinfo  {journal} {SIAM J. Optim.}\ }\textbf
  {\bibinfo {volume} {13}},\ \bibinfo {pages} {889} (\bibinfo {year}
  {2002})}\BibitemShut {NoStop}%
\bibitem [{\citenamefont {Dolan}\ \emph {et~al.}(2003)\citenamefont {Dolan},
  \citenamefont {Lewis},\ and\ \citenamefont {Torczon}}]{Dolan:2003}%
  \BibitemOpen
  \bibfield  {author} {\bibinfo {author} {\bibfnamefont {E.~D.}\ \bibnamefont
  {Dolan}}, \bibinfo {author} {\bibfnamefont {R.~M.}\ \bibnamefont {Lewis}}, \
  and\ \bibinfo {author} {\bibfnamefont {V.}~\bibnamefont {Torczon}},\
  }\bibfield  {title} {\emph {\bibinfo {title} {On the local convergence of
  pattern search}},\ }\href {\doibase 10.1137/S1052623400374495} {\bibfield
  {journal} {\bibinfo  {journal} {SIAM J. Optim.}\ }\textbf {\bibinfo {volume}
  {14}},\ \bibinfo {pages} {567} (\bibinfo {year} {2003})}\BibitemShut
  {NoStop}%
\bibitem [{\citenamefont {Lagarias}\ \emph {et~al.}(1998)\citenamefont
  {Lagarias}, \citenamefont {Reeds}, \citenamefont {Wright},\ and\
  \citenamefont {Wright}}]{Lagarias:1998}%
  \BibitemOpen
  \bibfield  {author} {\bibinfo {author} {\bibfnamefont {J.~C.}\ \bibnamefont
  {Lagarias}}, \bibinfo {author} {\bibfnamefont {J.~A.}\ \bibnamefont {Reeds}},
  \bibinfo {author} {\bibfnamefont {M.~H.}\ \bibnamefont {Wright}}, \ and\
  \bibinfo {author} {\bibfnamefont {P.~E.}\ \bibnamefont {Wright}},\ }\bibfield
   {title} {\emph {\bibinfo {title} {Convergence properties of the
  {Nelder-Mead} simplex method in low dimensions}},\ }\href {\doibase
  10.1137/S1052623496303470} {\bibfield  {journal} {\bibinfo  {journal} {SIAM
  J. Optim.}\ }\textbf {\bibinfo {volume} {9}},\ \bibinfo {pages} {112}
  (\bibinfo {year} {1998})}\BibitemShut {NoStop}%
\bibitem [{\citenamefont {McKinnon}(1998)}]{McKinnon:1998}%
  \BibitemOpen
  \bibfield  {author} {\bibinfo {author} {\bibfnamefont {K.~I.}\ \bibnamefont
  {McKinnon}},\ }\bibfield  {title} {\emph {\bibinfo {title} {Convergence of
  the {Nelder-Mead} simplex method to a nonstationary point}},\ }\href
  {\doibase 10.1137/S1052623496303482} {\bibfield  {journal} {\bibinfo
  {journal} {SIAM J. Optim.}\ }\textbf {\bibinfo {volume} {9}},\ \bibinfo
  {pages} {148} (\bibinfo {year} {1998})}\BibitemShut {NoStop}%
\bibitem [{\citenamefont {Rios}\ and\ \citenamefont
  {Sahinidis}(2013)}]{Rios:2013}%
  \BibitemOpen
  \bibfield  {author} {\bibinfo {author} {\bibfnamefont {L.~M.}\ \bibnamefont
  {Rios}}\ and\ \bibinfo {author} {\bibfnamefont {N.~V.}\ \bibnamefont
  {Sahinidis}},\ }\bibfield  {title} {\emph {\bibinfo {title} {Derivative-free
  optimization: a review of algorithms and comparison of software
  implementations}},\ }\href {\doibase 10.1007/s10898-012-9951-y} {\bibfield
  {journal} {\bibinfo  {journal} {J. Global Optim.}\ }\textbf {\bibinfo
  {volume} {56}},\ \bibinfo {pages} {1247} (\bibinfo {year}
  {2013})}\BibitemShut {NoStop}%
\bibitem [{\citenamefont {Loshchilov}\ \emph {et~al.}(2011)\citenamefont
  {Loshchilov}, \citenamefont {Schoenauer},\ and\ \citenamefont
  {Sebag}}]{Loshchilov:2011}%
  \BibitemOpen
  \bibfield  {author} {\bibinfo {author} {\bibfnamefont {I.}~\bibnamefont
  {Loshchilov}}, \bibinfo {author} {\bibfnamefont {M.}~\bibnamefont
  {Schoenauer}}, \ and\ \bibinfo {author} {\bibfnamefont {M.}~\bibnamefont
  {Sebag}},\ }\bibfield  {title} {\emph {\bibinfo {title} {Adaptive coordinate
  descent}},\ }in\ \href {\doibase 10.1145/2001576.2001697} {\emph {\bibinfo
  {booktitle} {Proc. of the 13th Annual Conference on Genetic and Evolutionary
  Computation}}}\ (\bibinfo {organization} {ACM},\ \bibinfo {year} {2011})\
  pp.\ \bibinfo {pages} {885--892}\BibitemShut {NoStop}%
\bibitem [{\citenamefont {Rinnooy~Kan}\ and\ \citenamefont
  {Timmer}(1987{\natexlab{a}})}]{RinnooyKan:1987a}%
  \BibitemOpen
  \bibfield  {author} {\bibinfo {author} {\bibfnamefont {A.~H.~G.}\
  \bibnamefont {Rinnooy~Kan}}\ and\ \bibinfo {author} {\bibfnamefont {G.~T.}\
  \bibnamefont {Timmer}},\ }\bibfield  {title} {\emph {\bibinfo {title}
  {Stochastic global optimization methods. {Part I}: Clustering methods}},\
  }\href {\doibase 10.1007/BF02592070} {\bibfield  {journal} {\bibinfo
  {journal} {Math. Prog.}\ }\textbf {\bibinfo {volume} {39}},\ \bibinfo {pages}
  {27} (\bibinfo {year} {1987}{\natexlab{a}})}\BibitemShut {NoStop}%
\bibitem [{\citenamefont {Rinnooy~Kan}\ and\ \citenamefont
  {Timmer}(1987{\natexlab{b}})}]{RinnooyKan:1987b}%
  \BibitemOpen
  \bibfield  {author} {\bibinfo {author} {\bibfnamefont {A.~H.~G.}\
  \bibnamefont {Rinnooy~Kan}}\ and\ \bibinfo {author} {\bibfnamefont {G.~T.}\
  \bibnamefont {Timmer}},\ }\bibfield  {title} {\emph {\bibinfo {title}
  {Stochastic global optimization methods. {Part II}: Multi level methods}},\
  }\href {\doibase 10.1007/BF02592071} {\bibfield  {journal} {\bibinfo
  {journal} {Math. Prog.}\ }\textbf {\bibinfo {volume} {39}},\ \bibinfo {pages}
  {57} (\bibinfo {year} {1987}{\natexlab{b}})}\BibitemShut {NoStop}%
\bibitem [{\citenamefont {Kennedy}(2011)}]{Kennedy:2011}%
  \BibitemOpen
  \bibfield  {author} {\bibinfo {author} {\bibfnamefont {J.}~\bibnamefont
  {Kennedy}},\ }\bibfield  {title} {\emph {\bibinfo {title} {Particle swarm
  optimization}},\ }in\ \href {\doibase 10.1007/978-0-387-30164-8_630} {\emph
  {\bibinfo {booktitle} {Encyclopedia of Machine Learning}}}\ (\bibinfo
  {publisher} {Springer},\ \bibinfo {year} {2011})\ pp.\ \bibinfo {pages}
  {760--766}\BibitemShut {NoStop}%
\bibitem [{\citenamefont {Poli}\ \emph {et~al.}(2007)\citenamefont {Poli},
  \citenamefont {Kennedy},\ and\ \citenamefont {Blackwell}}]{Poli:2007}%
  \BibitemOpen
  \bibfield  {author} {\bibinfo {author} {\bibfnamefont {R.}~\bibnamefont
  {Poli}}, \bibinfo {author} {\bibfnamefont {J.}~\bibnamefont {Kennedy}}, \
  and\ \bibinfo {author} {\bibfnamefont {T.}~\bibnamefont {Blackwell}},\
  }\bibfield  {title} {\emph {\bibinfo {title} {Particle swarm optimization}},\
  }\href {\doibase 10.1007/s11721-007-0002-0} {\bibfield  {journal} {\bibinfo
  {journal} {Swarm Intell.}\ }\textbf {\bibinfo {volume} {1}},\ \bibinfo
  {pages} {33} (\bibinfo {year} {2007})}\BibitemShut {NoStop}%
\bibitem [{\citenamefont {Trelea}(2003)}]{Trelea:2003}%
  \BibitemOpen
  \bibfield  {author} {\bibinfo {author} {\bibfnamefont {I.~C.}\ \bibnamefont
  {Trelea}},\ }\bibfield  {title} {\emph {\bibinfo {title} {The particle swarm
  optimization algorithm: convergence analysis and parameter selection}},\
  }\href {\doibase 10.1016/S0020-0190(02)00447-7} {\bibfield  {journal}
  {\bibinfo  {journal} {Inform. Process. Lett.}\ }\textbf {\bibinfo {volume}
  {85}},\ \bibinfo {pages} {317} (\bibinfo {year} {2003})}\BibitemShut
  {NoStop}%
\bibitem [{\citenamefont {Powell}(1973)}]{Powell:1973}%
  \BibitemOpen
  \bibfield  {author} {\bibinfo {author} {\bibfnamefont {M.~J.~D.}\
  \bibnamefont {Powell}},\ }\bibfield  {title} {\emph {\bibinfo {title} {On
  search directions for minimization algorithms}},\ }\href {\doibase
  10.1007/BF01584660} {\bibfield  {journal} {\bibinfo  {journal} {Math. Prog.}\
  }\textbf {\bibinfo {volume} {4}},\ \bibinfo {pages} {193} (\bibinfo {year}
  {1973})}\BibitemShut {NoStop}%
\bibitem [{\citenamefont {Price}\ \emph {et~al.}(2002)\citenamefont {Price},
  \citenamefont {Coope},\ and\ \citenamefont {Byatt}}]{Price:2002}%
  \BibitemOpen
  \bibfield  {author} {\bibinfo {author} {\bibfnamefont {C.}~\bibnamefont
  {Price}}, \bibinfo {author} {\bibfnamefont {I.}~\bibnamefont {Coope}}, \ and\
  \bibinfo {author} {\bibfnamefont {D.}~\bibnamefont {Byatt}},\ }\bibfield
  {title} {\emph {\bibinfo {title} {A convergent variant of the {Nelder-Mead}
  algorithm}},\ }\href {\doibase 10.1023/A:1014849028575} {\bibfield  {journal}
  {\bibinfo  {journal} {J. Optim. Theory Appl.}\ }\textbf {\bibinfo {volume}
  {113}},\ \bibinfo {pages} {5} (\bibinfo {year} {2002})}\BibitemShut {NoStop}%
\bibitem [{\citenamefont {Nazareth}\ and\ \citenamefont
  {Tseng}(2002)}]{Nazareth:2002}%
  \BibitemOpen
  \bibfield  {author} {\bibinfo {author} {\bibfnamefont {L.}~\bibnamefont
  {Nazareth}}\ and\ \bibinfo {author} {\bibfnamefont {P.}~\bibnamefont
  {Tseng}},\ }\bibfield  {title} {\emph {\bibinfo {title} {Gilding the {Lily}:
  a variant of the {Nelder-Mead} algorithm based on golden-section search}},\
  }\href {\doibase 10.1023/A:1014842520519} {\bibfield  {journal} {\bibinfo
  {journal} {Comp. Optim. Appl.}\ }\textbf {\bibinfo {volume} {22}},\ \bibinfo
  {pages} {133} (\bibinfo {year} {2002})}\BibitemShut {NoStop}%
\bibitem [{\citenamefont {Conn}\ \emph {et~al.}(2009)\citenamefont {Conn},
  \citenamefont {Scheinberg},\ and\ \citenamefont {Vicente}}]{Conn:2009}%
  \BibitemOpen
  \bibfield  {author} {\bibinfo {author} {\bibfnamefont {A.~R.}\ \bibnamefont
  {Conn}}, \bibinfo {author} {\bibfnamefont {K.}~\bibnamefont {Scheinberg}}, \
  and\ \bibinfo {author} {\bibfnamefont {L.~N.}\ \bibnamefont {Vicente}},\
  }\emph {\bibinfo {title} {Simplicial direct-search methods}},\ in\ \href
  {\doibase 10.1137/1.9780898718768.ch8} {\emph {\bibinfo {booktitle}
  {Introduction to derivative-free optimization}}}\ (\bibinfo  {publisher}
  {SIAM},\ \bibinfo {year} {2009})\ Chap.~\bibinfo {chapter} {8}, pp.\ \bibinfo
  {pages} {141--162}\BibitemShut {NoStop}%
\bibitem [{\citenamefont {Koshelev}\ \emph {et~al.}(2016)\citenamefont
  {Koshelev}, \citenamefont {Sadovskyy}, \citenamefont {Phillips},\ and\
  \citenamefont {Glatz}}]{Koshelev:2016}%
  \BibitemOpen
  \bibfield  {author} {\bibinfo {author} {\bibfnamefont {A.~E.}\ \bibnamefont
  {Koshelev}}, \bibinfo {author} {\bibfnamefont {I.~A.}\ \bibnamefont
  {Sadovskyy}}, \bibinfo {author} {\bibfnamefont {C.~L.}\ \bibnamefont
  {Phillips}}, \ and\ \bibinfo {author} {\bibfnamefont {A.}~\bibnamefont
  {Glatz}},\ }\bibfield  {title} {\emph {\bibinfo {title} {Optimization of
  vortex pinning by nanoparticles using simulations of the time-dependent
  {Ginzburg-Landau} model}},\ }\href {\doibase 10.1103/PhysRevB.93.060508}
  {\bibfield  {journal} {\bibinfo  {journal} {Phys. Rev. B}\ }\textbf {\bibinfo
  {volume} {93}},\ \bibinfo {pages} {060508} (\bibinfo {year}
  {2016})}\BibitemShut {NoStop}%
\end{thebibliography}

%


\end{document}